\documentclass[letter]{aa}  
\usepackage{graphicx}
\usepackage{txfonts}

\begin{document} 

\title{A high pitch angle structure in the Sagittarius Arm}

   \author{M. A. Kuhn
          \inst{1}
          \and
          R. A. Benjamin\inst{2}
           \and
          C. Zucker\inst{3}
           \and
          A. Krone-Martins\inst{4,5}
           \and
          R. S. de Souza\inst{6}
           \and
          A. Castro-Ginard\inst{7}
           \and
          E. E. O. Ishida\inst{8}
           \and
          M. S. Povich\inst{9}
           \and
          L. A. Hillenbrand\inst{1}
          for the COIN Collaboration
          }

   \institute{California Institute of Technology, Pasadena, CA 91125, USA
              (\email{mkuhn@astro.caltech.edu})
              \and
             Department of Physics, University of Wisconsin-Whitewater, 800 W Main St, Whitewater, WI 53190 USA
                      \and
           Center for Astrophysics $\vert$ Harvard \& Smithsonian, 60 Garden St, Cambridge, MA, 02138 USA
                      \and
             Donald Bren School of Information and Computer Sciences, University of California, Irvine, CA 92697, USA
                      \and
             CENTRA/SIM, Faculdade de Ciências, Universidade de Lisboa, Ed. C8, Campo Grande, 1749-016, Lisboa, Portugal
                      \and
             Key Laboratory for Research in Galaxies and Cosmology, Shanghai Astronomical Observatory,
Chinese Academy of Sciences, 80 Nandan Rd., Shanghai 200030, China
                      \and
             Institut de Ciències del Cosmos, Universitat de Barcelona (IEEC-UB), Martí i Franquès 1, 08028 Barcelona, Spain
                      \and
             Université Clermont Auvergne, CNRS/IN2P3, LPC, F-63000 Clermont-Ferrand, France
                      \and
             Department of Physics and Astronomy, California State Polytechnic University Pomona, 3801 West Temple Avenue, Pomona, CA
91768, USA
\\
             }
             
   \date{Received April 27, 2021; Accepted June 28, 2021}

  \abstract
   {In spiral galaxies, star formation tends to trace features of the spiral pattern, including arms, spurs, feathers, and branches. However, in our own Milky Way, it has been challenging to connect individual star-forming regions to their larger Galactic environment owing to our perspective from within the disk. One feature in nearly all modern models of the Milky Way is the Sagittarius Arm, located inward of the Sun with a pitch angle of $\sim$12$^{\circ}$. 
   }
   {We map the 3D locations and velocities of star-forming regions in a segment of the Sagittarius Arm using young stellar objects (YSOs) from the {\it Spitzer}/IRAC Candidate YSO (SPICY) catalog to compare their distribution to models of the arm.}
  {Distances and velocities for these objects are derived from Gaia EDR3 astrometry and molecular line surveys. We infer parallaxes and proper motions for spatially clustered groups of YSOs and estimate their radial velocities from the velocities of spatially associated molecular clouds. }
   {We identify 25 star-forming regions in the Galactic longitude range $\ell\sim4.\!{^\circ}0$--$18.\!{^\circ}5$ arranged in a narrow, $\sim$1 kpc long linear structure with a high pitch angle of $\psi = 56^\circ$ and a high aspect ratio of $\sim$7:1.
   This structure includes massive star-forming regions such as M8, M16, M17, and M20. The motions in the structure are remarkably coherent, with velocities in the direction of Galactic rotation of $|V_\varphi| \approx 240\pm3$~km~s$^{-1}$ (slightly higher than average) and slight drifts inward ($V_\mathrm{R} \approx -4.3$~km~s$^{-1}$) and in the negative $Z$ direction ($V_\mathrm{Z} \approx -2.9$~km~s$^{-1}$). The rotational shear experienced by the structure is $\Delta\Omega=4.6~$km~s$^{-1}$~kpc$^{-1}$.}
   {The observed $56^\circ$ pitch angle is remarkably high for a segment of the Sagittarius Arm. We discuss possible interpretations of this feature as a substructure within the lower pitch angle Sagittarius Arm, as a spur, or as an isolated structure.}
   \keywords{Galaxy: structure -- Galaxy: kinematics and dynamics -- Galaxies: spiral -- ISM: clouds --Stars: formation}

  \maketitle

\section{Introduction}

Most of our understanding of spiral arms comes from observations of other galaxies, where our outside perspective allows us to see the full spiral structure \citep{Sandage1961,vanDerKruit2011}. In these other galaxies, spiral arms often have smaller-scale structures, including spurs (luminous stellar features) and feathers (dust features) that extend from arms to inter-arm regions, as well as branches in the main arms \citep[][]{Elmegreen1980,LaVigne2006}. In the Milky Way, it has been more challenging to disentangle such features owing to our perspective within the highly extincted disk. 

In the current picture of the Milky Way, the Sagittarius Arm is the closest major spiral arm inward from the Sun and hosts several prominent, nearby massive star-forming regions. Early-type stars in the regions M8, M16, M20, and several others, were used by \citet{Morgan1953} to define the Sagittarius Arm in the first widely accepted Galactic map to show spiral structure (Appendix~\ref{sec:history}). The currently favored four-armed model is largely based on H\,{\sc i} and CO emission in longitude-velocity ($\ell$--$v$) diagrams \citep[e.g.,][]{Dame2001} supplemented with very long baseline interferometric (VLBI) parallax measurements of masers \cite[e.g., BeSSeL and VERA;][]{Reid2019,Hirota_2020}. Conversion of the $\ell$--$v$ diagram to face-on maps has generally required assumptions regarding circular orbits. Although the maser sample has refined this approach, the number of masing targets is small compared to the number of star-forming regions and, furthermore, currently limited to Northern Hemisphere targets.

\begin{figure*}[t]
        \centering
        \includegraphics[width=0.900\textwidth]{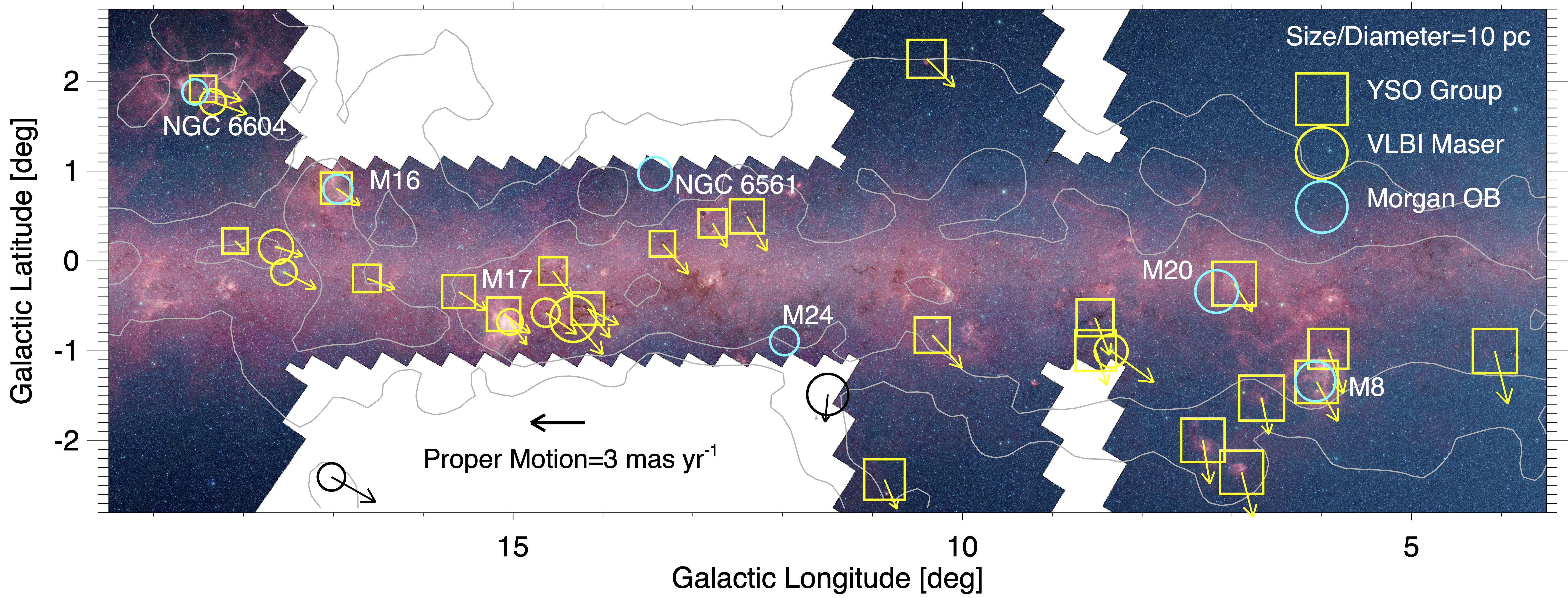}
    \caption{False-color GLIMPSE survey mosaic \citep{Benjamin2003,Churchwell2009} of the Sagittarius Arm structure. 
    The three white contours show the intensity of $^{12}$CO emission (10, 31, and 100~K~km~s$^{-1}$) from \citet{Dame2001} integrated over the velocity range $v_\mathrm{lsr}=5$--30~km~s$^{-1}$. Symbols mark the YSO groups, masers, and the Morgan OB associations (reanalyzed with \textit{Gaia}) that make up the structure. Smaller boxes and circles indicate greater distances, and arrows indicate proper motions.}
  \label{fig:GLIMPSE}
\end{figure*}

\begin{figure}
        \centering
        \includegraphics[width=0.47\textwidth]{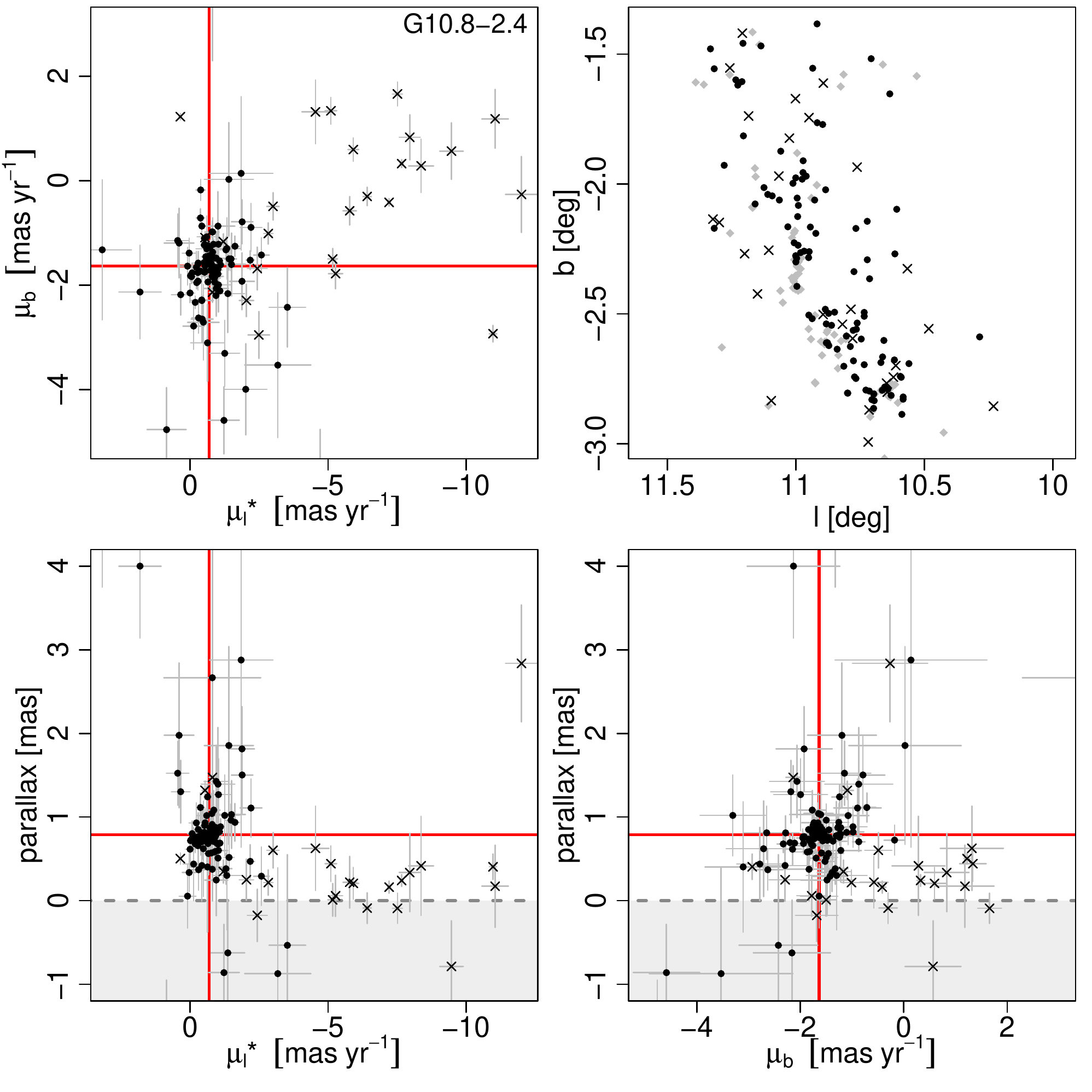}
    \caption{Scatter plots of stellar astrometry for one of the YSO groups. Stars identified by the Bayesian model as probable members are black circles, probable contaminants are indicated with black ``x'' marks, and stars without Gaia astrometry are gray diamonds. 1$\sigma$ uncertainties are indicated by the error bars. The red lines show the group's mean parallax and proper motion, with 95\% credible intervals indicated by shading around the lines. (Here, the results are precise enough that the shaded area is difficult to discern.) Negative distances (gray region) are excluded by the prior.}
  \label{fig:clust_model}
\end{figure}

The availability of parallaxes and proper motions for over a billion sources \citep{Gaia2016,Gaia2018,Gaia2020} has led to a renaissance in investigations of Galactic spiral structure within a few kiloparsecs of the Sun. \citet{Zucker2020} and \citet{Alves2020} found a coherent, filamentary structure formed by star-forming clouds, likely associated with the Local Arm.   
\citet{Xu2021} and \citet{PantaleoniGonzalez2021} examined the spatial distribution of previously identified OB stars, while \citet{Zari2021} and \citet{Poggio2021} have characterized a photometrically selected sample of upper-main-sequence stars. 

Here, we trace Galactic structure using young stellar objects (YSOs), which provide a link between the stellar content of spiral arms and the molecular clouds in  which the YSOs form. This letter focuses on a distinct linear feature between Galactic longitudes $\ell\approx 4^\circ$--$18.\!{^\circ}5$ (Fig.~\ref{fig:GLIMPSE}). In this region, the suggestion of a linear structure is visible in Galactic maps produced by several previous studies, including the 3D extinction maps from \citet{Green2019}, the cloud distances from \citet{Zucker2020}, the maser distances from \citet{Reid2019}, and the Gaia Data Release 2 (DR2) cluster distances from \citet{spicyI}, and is even hinted at in the original \citet{Morgan1953} association distances. However, the discrepancy between the high pitch angle of the structure we trace here and pitch angles used in models of the Sagittarius Arm has never before, to our knowledge, been remarked upon. 

\section{Astrometry for YSO groups}

\subsection{Candidate YSOs in Gaia EDR3}

\defcitealias{spicyI}{Paper~I}

% This is the most accurate map to-date of the territory of the Ferengi Alliance. 
\begin{figure*}
        \centering
        \includegraphics[width=0.47\textwidth]{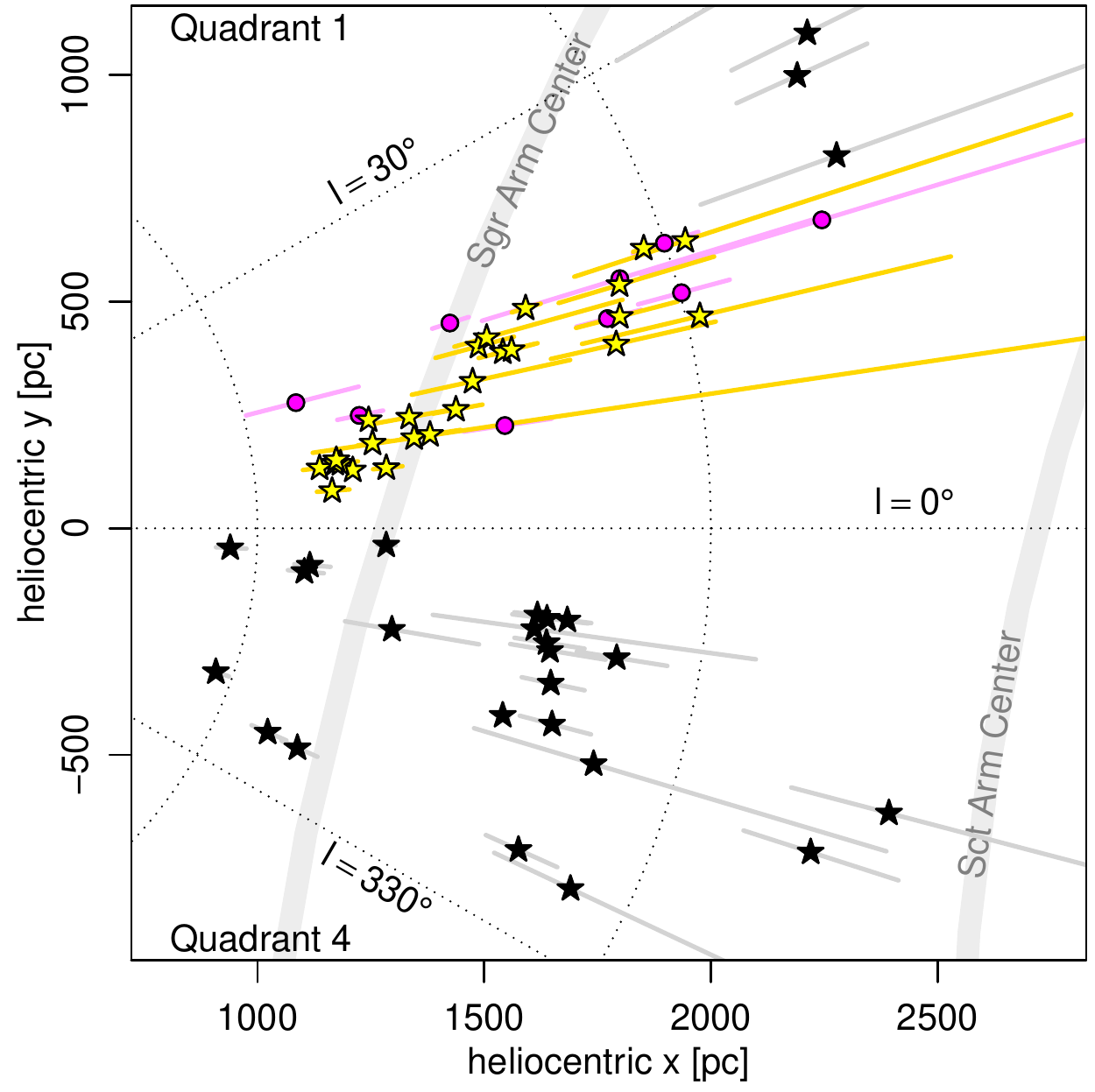}
        \includegraphics[width=0.47\textwidth]{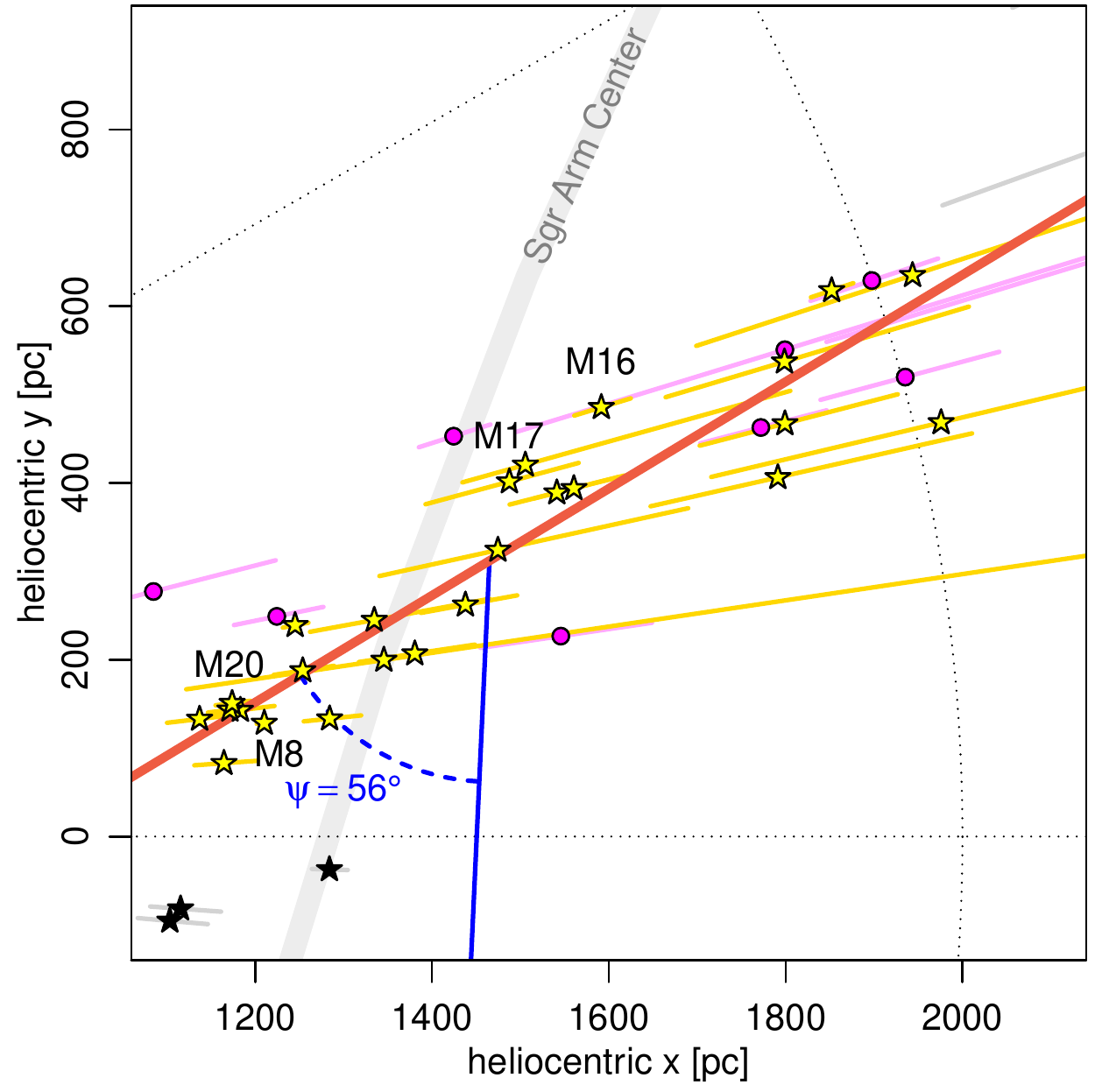}
    \caption{Map of YSO groups (star symbols) and masers (magenta circles) in heliocentric coordinates. The right panel shows a zoomed-in view; it should be noted that the Sun is off the left side of both plots. Groups associated with the structure are highlighted in yellow, while others are in black. The spiral-arm centers defined by \citet{Reid2019} are indicated by the gray bands. The red line indicates the major axis of the feature identified here, with its 56$^\circ$ pitch angle illustrated in blue.
    }
  \label{fig:heliocentric_distribution}
\end{figure*}

\begin{figure}
        \centering
        \includegraphics[width=0.45\textwidth]{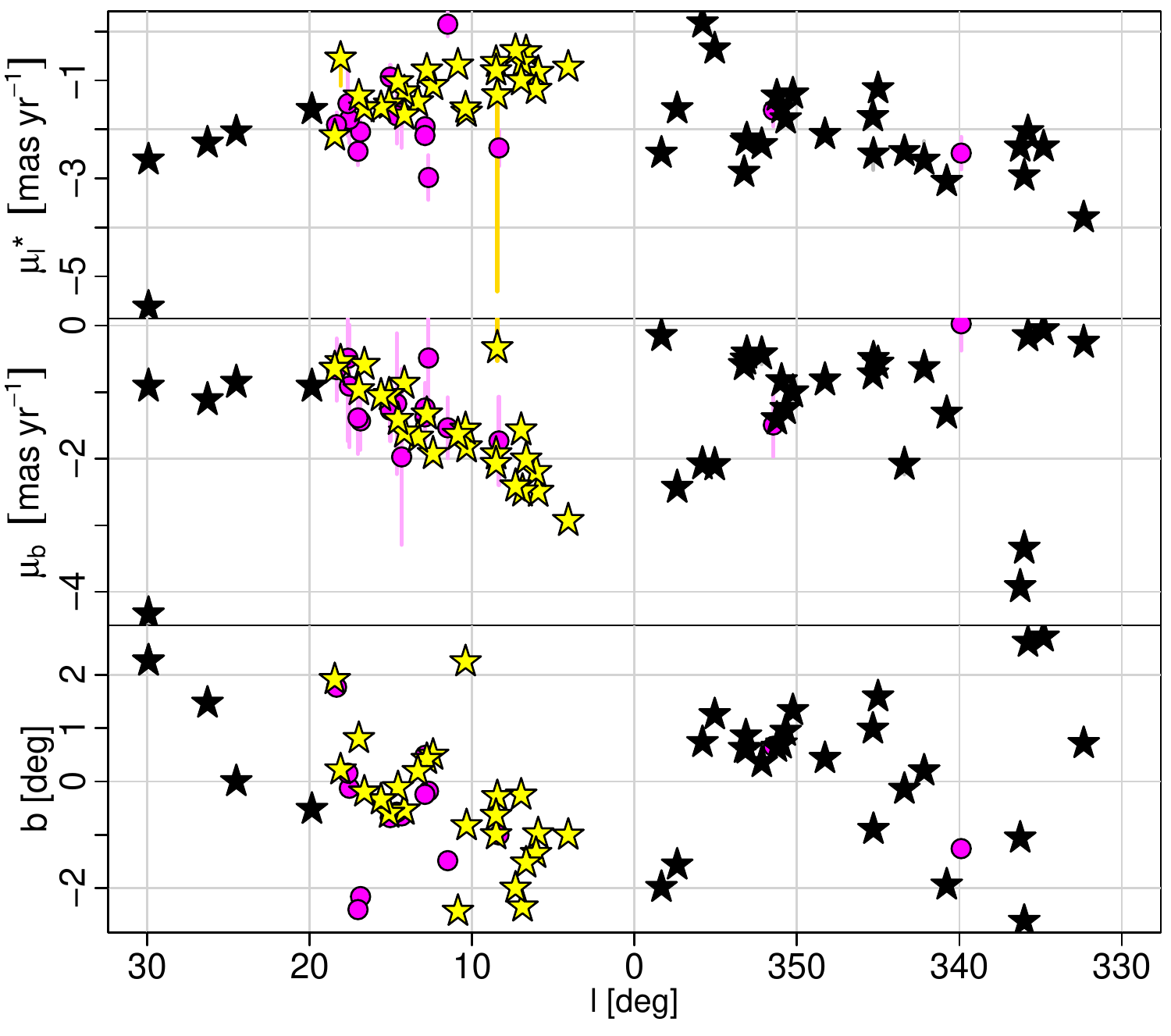}
    \vspace*{-0.05in}
    \caption{Astrometric quantities ($\mu_{\ell^\star}$, $\mu_b$, and $b$) versus Galactic longitude $\ell$ for YSO groups and masers (same symbols as Fig.~\ref{fig:heliocentric_distribution}). The missing error bars are smaller than the symbols.}
  \label{fig:angular_distribution}
\end{figure}

Our study is based on stars from the {\it Spitzer}/IRAC Candidate YSO  (SPICY)  catalog \citep[][hereafter Paper~I]{spicyI}. These objects were identified via mid-infrared indications of circumstellar disks or envelopes, based on photometry from large {\it Spitzer} surveys of the Galactic midplane, including GLIMPSE \citep{Benjamin2003,Churchwell2009} and related surveys. The full catalog contains 117,446 YSO candidates, covering nearly all of the first and fourth Galactic quadrants between $|b|\lesssim1$--2$^\circ$. However, in this letter we analyze objects between $-30^\circ\leq\ell<30^\circ$.

\citetalias{spicyI} divides the YSOs into groups based on clustering in $(\ell, b)$ coordinates. These groups were defined by the HDBSCAN algorithm \citep{Campello2013} requiring $\geq$30 stars per group. From these criteria, half the YSO candidates are members of groups.

Roughly one-third of the YSO candidates are matched to Gaia Early Data Release 3 (EDR3) counterparts within a radius of 1$^{\prime\prime}$.\footnote{This radius is selected to be several times the typical absolute astrometric accuracy of the GLIMPSE catalog. To estimate the spurious match rate, we shifted right ascensions by 5$^\prime$ and obtained a 3\% match rate. For sources with matches to both the true and shifted positions, the separations were smaller for the true position 90\% of the time. This suggests that that spurious match rate is small and that true matches will usually override spurious matches. %Remaining spurious matches will generally not have the same parallax and proper motion, so will not be considered likely cluster members.
}
We used only stars with renormalized unit weight errors of $\leq$1.4 \citep{Lindegren2020_as} and applied the \citet{Lindegren2020_zp} parallax zero-point correction, which is estimated as a function of color, magnitude, and ecliptic latitude. These corrections may be especially useful for YSOs in regions with significant differential absorption that can cause large color differences between objects in the same cluster \citep[e.g.,][]{KuhnHillenbrand2020}.

\subsection{Estimating mean parallaxes and proper motions for groups}
\label{sec:means}

Parallaxes, $\varpi$, and proper motions in Galactic longitude, $\mu_{\ell^\star} = \mu_{\ell}\cos(b)$, and latitude, $\mu_b$, can often be more accurately determined for groups of stars than for individual stars. However, estimates of mean astrometric properties may be affected by heteroscedastic uncertainties, velocity dispersions within groups, and contaminants (both YSOs at different distances and non-YSOs).

To take full advantage of the improved Gaia~EDR3 astrometry, we employed a Bayesian model of the astrometric measurements for stars in each YSO group. Outliers can have high leverage on means, so we included a mixture component in the model to account for contaminants \citep[e.g.,][]{Hogg2010,HoMA}. {\it Bona fide} members are expected to cluster in parallax and proper-motion space, but nonmembers tend to have broader distributions that depend on $\ell$ and $b$. This model, the details of which are provided in Appendices~\ref{sec:model} and \ref{sec:non-yso}, is fit via Markov chain Monte Carlo (MCMC). 

Figure~\ref{fig:clust_model} shows stars from one of the YSO groups plotted in parallax and proper-motion space and fit with this model. The probable members form a distinct cluster in this space, while probable nonmembers have a wider range of values. For the subsequent analysis, we used groups with at least eight probable members to ensure that a reliable cluster exists.\footnote{The results are not particularly sensitive to this threshold; for example, requiring $\geq$2 members only yields $\sim$3 more groups that appear associated with the feature analyzed here.} In general, the uncertainties on the mean parallaxes and proper motions of the groups tend to be smaller than those of the individual stars.

\subsection{A kiloparsec-long structure in the Sagittarius Arm}

\begin{figure*}
        \centering
        \includegraphics[width=1\textwidth]{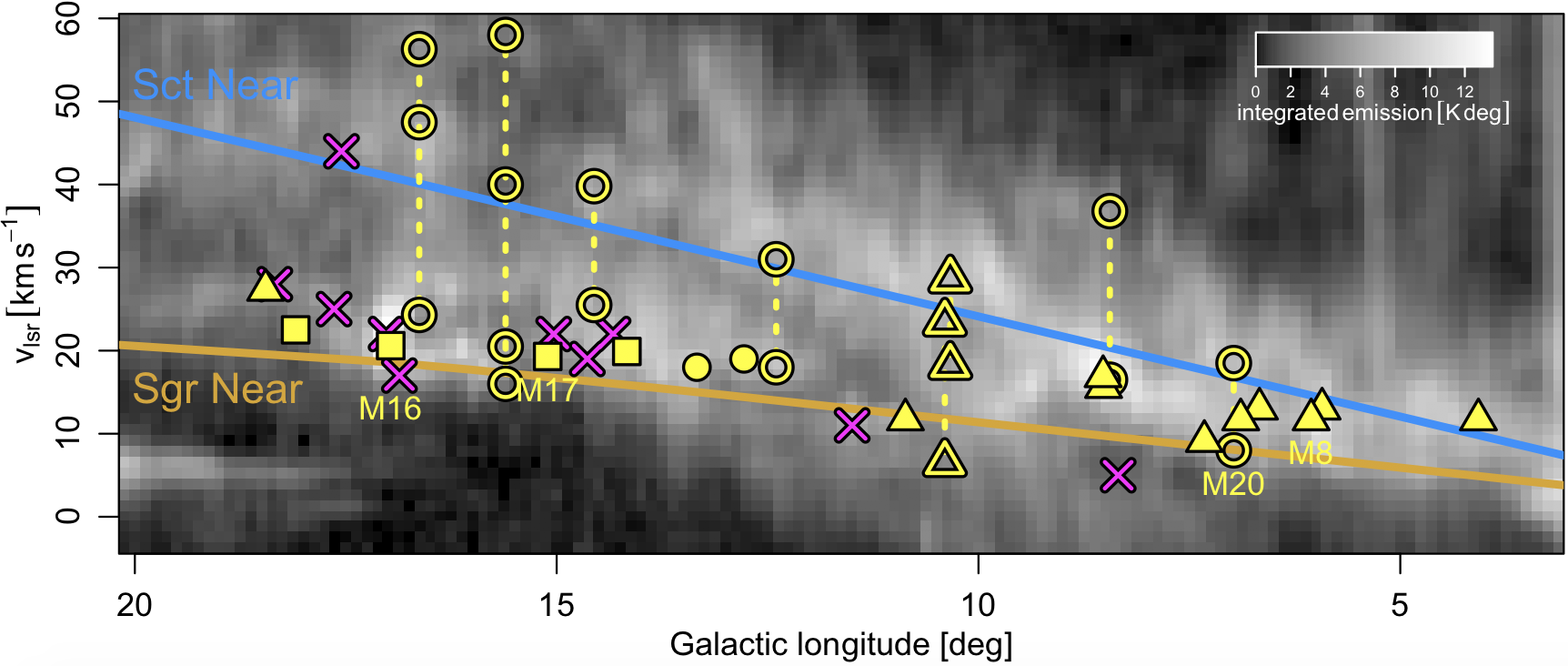}
    \caption{ $\ell$--$v$ diagram for objects in the new structure. Groups associated with single velocity peaks are filled symbols, while ambiguous cases are open symbols with dashed lines connecting plausible $v_\mathrm{lsr}$ solutions. The shape indicates the survey used to determine the $v_\mathrm{lsr}$: SEDIGISM (circle), FUGIN (square), and COGAL (triangle). Masers are plotted as magenta ``x'' marks. The underlying image is the \citet{Dame2001} COGAL map integrated over $|b|\leq3^\circ$, with the loci of the Sagittarius Arm (orange line) and Scutum Arm (blue line) from \citet{Reid2016}. Each YSO group in the structure has a possible solution consistent with the Sagittarius Arm.}
  \label{fig:COGAL_arm}
\end{figure*}

The 3D positions of the YSO groups, based on the group parallaxes, form an elongated linear feature in the first Galactic quadrant. In Fig.~\ref{fig:heliocentric_distribution}, the YSO groups (star symbols) are plotted in a heliocentric Cartesian coordinate system, where the Galactic center is on the positive $x$ axis, the $y$ axis is parallel to the direction of Galactic rotation, and the $z$ axis points out of the plane following a right-handed system (Appendix~\ref{sec:table}). The YSO groups making up the structure (highlighted in yellow) range from $\sim$1--2~kpc in heliocentric distance, form a remarkably narrow band angled relative to our line of sight, and follow coherent patterns in both proper motion (Fig.~\ref{fig:angular_distribution}) and $v_\mathrm{lsr}$ (Fig.~\ref{fig:COGAL_arm}). 

Several prominent massive star-forming regions are among the 25 YSO groups defining the structure. These include M8 (Lagoon), M16 (Eagle), M17 (Omega), M20 (Trifid), NGC 6559, and Sharpless 54, but there are also many smaller YSO groups without ionizing stars. Furthermore, ten masers associated with massive star formation or red supergiant stars from the \citet{Reid2019} sample (magenta circles in Figs.~\ref{fig:heliocentric_distribution} and \ref{fig:angular_distribution}) are aligned with this structure. Finally, these objects lie within a filamentary region of high dust extinction seen in 3D dust maps (Appendix~\ref{sec:av}).
\textit{Gaia's} ability to constrain distances declines beyond a few kiloparsecs, so it is possible that the structure could extend farther inward in galactocentric radius than we can detect from the EDR3 data. 

A notable property of the new structure is its high pitch angle relative to the commonly assumed low pitch angle of the Sagittarius Arm. The structure is centered at $(x,y,z)=(1470\pm50, 310\pm30, -8\pm6)$~pc. The principal axis of the structure is parallel to  $(0.85\pm0.02)\,\hat{x}+(0.52\pm0.04)\,\hat{y} + (0.05\pm0.02)\,\hat{z}$. The YSO groups extend $\sim$950~pc along this axis, and the aspect ratio of the structure is $\sim$14:2:1, with the structure being narrowest in the $\hat{z}$ direction. Based on this orientation, the structure has a pitch angle $\psi=56^\circ$.

Additional evidence for the coherence of this structure comes from kinematics. In proper motion versus Galactic longitude (Fig.~\ref{fig:angular_distribution}), the objects forming the structure are tightly aligned in narrow bands; masers confirm these trends. In $\mu_{\ell^\star}$ versus\ $\ell$, its groups approximately follow the same curve as other YSO groups and masers. And in $\mu_b$ versus\ $\ell$, the structure is also coherent, with the near end having more negative $\mu_b$ values. Furthermore, the near end of the structure has lower average $b$. These trends can be partially attributed to perspective effects of the Sun's position above the Galactic plane with a positive vertical velocity \citep{Bland-Hawthorn_Gerhard2016,Reid2019}. 

\begin{figure*}
        \centering
                \includegraphics[width=0.95\textwidth]{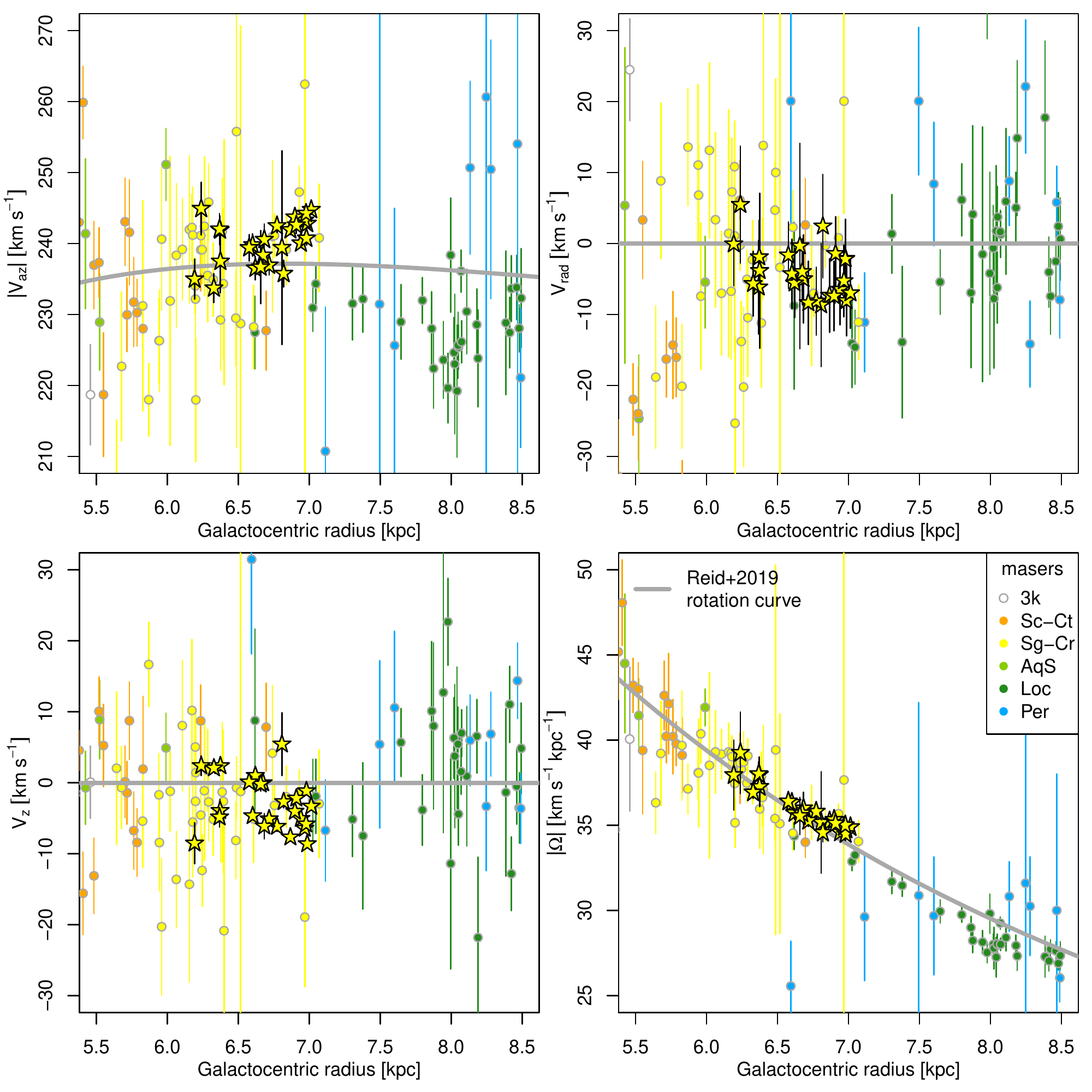}
    \caption{Rotational velocity, radial velocity, vertical velocity, and angular velocity versus\ galactocentric radius. YSO groups (star symbols) and masers (circles) are color-coded based on the spiral arm categorizations from \citet[][see abbreviations therein]{Reid2019}. The gray lines indicate their rotation curve. }
  \label{fig:galactocentric}
\end{figure*}

Finally, to estimate radial velocities of the YSO groups, we associated them with molecular gas observed in CO surveys of the Galactic plane (Appendix~\ref{sec:CO}), including SEDIGISM \citep[][]{Schuller2021}, FUGIN \citep[][]{Umemoto2017}, and COGAL data \citep[][]{Dame2001}. On the $\ell$--$v$ diagram (Fig.~\ref{fig:COGAL_arm}), the groups with known velocities range from $v_\mathrm{lsr}\approx 10$--30~km~s$^{-1}$, with a gradual increase with increasing $\ell$, and all YSO groups with uncertain velocities have possible solutions consistent with this trend. The YSO groups that we considered part of this new structure roughly follow the band in the COGAL data thought to be associated with the near Sagittarius Arm.\footnote{The maser with a discrepant $v_\mathrm{lsr}$ of 44~km~s$^{-1}$ is associated with the runaway red supergiant star IRC -10414 \citep{Gvaramadze2014}.}

\section{Kinematics of the structure}

To examine the spatial kinematics of the structure, we converted the observed coordinates into a cylindrical galactocentric coordinate system $(R, \varphi, Z)$ using a reference frame defined by the constants from \citet{Reid2019} (Appendix~\ref{sec:table}). In this system the galactocentric radii of the groups range from $R = 6.2$--7.0~kpc, and their vertical range is $Z = -50$--70~pc, with a mean of $Z = -2$~pc.

Trends in velocity with radius are shown in Fig.~\ref{fig:galactocentric}. The azimuthal velocities are $|V_\varphi| \sim 240\pm 3$~km~s$^{-1}$ -- slightly faster than the expected Galactic rotation -- with an increase in velocity with radius that is moderately statistically significant ($p < 0.01$ using Kendall's $\tau$ test). Most groups have a negative galactocentric radial velocity ($V_R = -4.3\pm3.5$~km~s$^{-1}$). The mean $V_Z$ is $-2.9$~km~s$^{-1}$, with a velocity dispersion of $\sim$4~km~s$^{-1}$. Finally, objects farther from the Galactic center have lower angular velocities than those closer ($p<10^{-3}$), indicating that the structure experiences shear of $\Delta\Omega = 4.6$~km~s$^{-1}$~kpc$^{-1}$. For comparison, Fig.~\ref{fig:galactocentric} also shows the masers (circles) that were used to establish the \citet{Reid2019} rotation curve.

Overall, these results suggest that the YSO groups in the feature have a coherent velocity structure, which is only $\sim$7~km~s$^{-1}$ discrepant from expectations for a purely circular orbit. This coherence provides further evidence that these groups are associated with one another rather than being a coincidental spatial arrangement. The presence of shear would explain the feature's trailing Galactic orientation. The observed shear implies a timescale of $\sim$90~Myr for the structure to change pitch angle by a factor of two \citep{Elmegreen1980}. Ages of young stars within the structure's star-forming regions are typically much younger than this \citep[e.g., $\sim$1 Myr;][]{Rho2008,Getman2014,Prisinzano2019}, but it remains to be determined whether any of the $\sim$12 open clusters in the vicinity of the structure with ages 10-100~Myr \citep{Cantat-Gaudin2020} are connected to it.

\section{Discussion and conclusion}

We have demonstrated that many of the first quadrant star-forming regions that, historically, have been used to define the Sagittarius Arm are part of an elongated structure with a pitch angle $\psi = 56^\circ$. This is higher than any pitch angle previously proposed for any portion the Sagittarius Arm (Appendix~\ref{sec:history}). This structure is traced by star-forming regions, masers, and 3D dust maps. Its distinct nature has been hiding in plain sight since the study of OB associations by \citet{Morgan1953}. But thanks to the dense and clustered sample of infrared-selected YSOs from the {\it Spitzer}/GLIMPSE program and the direct distance measurements from \textit{Gaia}, we can be confident that this is a real Galactic feature.

To envision how this structure would appear if viewed from outside the Milky Way, we can compare its star-formation activity to structures observed in other galaxies. {\it Spitzer}'s 24~$\mu$m band is a useful tracer of star formation since roughly 25\% of the bolometric thermal luminosity of star-forming regions is emitted in this band \citep{Binder2018}. The mid-infrared flux of our feature is dominated by its most significant star-forming regions (e.g., M8, M16, M17, and M20), which have a combined 24~$\mu$m luminosity of $8\times10^{39}$~erg~s$^{-1}$ \citep{Binder2018}. With a length of $\sim$950~pc and a width of $\sim$160~pc, the mean 24~$\mu$m surface brightness of the structure would be $\sim5\times10^{40}$~erg~s$^{-1}$~kpc$^{-2}$. This is near the middle of the \citet{Calzetti2007} sample of H\,{\sc ii} knots in nearby galaxies, suggesting that the structure would appear as a bright stellar feature. 

Although this structure is remarkable in the context of Milky Way models, numerous galaxies contain high pitch angle structures, some related to an overall spiral pattern and some whose origin is not so clear. For example, spurs and feathers in other galaxies have pitch angles ranging from $\sim$40--80$^\circ$ (mean of $\sim$60$^\circ$) and lengths ranging from 1--5~kpc \citep{Elmegreen1980,LaVigne2006} -- similar to the properties of the structure we have examined here. These structures extend from spiral arms to inter-arm regions, and they typically exhibit quasi-regular spacing with separations from $\sim$300--800~pc \citep{LaVigne2006}. Several theoretical models have been developed to explain the formation of spur-like structures in gaseous galactic disks, including formation due to gravitational instabilities and shear \citep{Balbus1988,Kim2002} with magnetohydrodynamical effects explored by \citet{Shetty2006}, formation due to hydrodynamics in spiral shocks \citep{Wada2004,Dobbs2006}, or expanding superbubbles \citep{Kim2020}. In the gravitational instability models, mass condensations form within the spiral arms, which are then sheared into the inter-arm regions to form spurs. In these models, even within the arm, the mass condensations are elongated with high pitch angles and lengths of $\sim$1~kpc. Thus, it is plausible that the feature we examined here corresponds to one of these mass concentrations within the Sagittarius Arm. 

Clarifying whether this feature is ({\it i}) an isolated structure, ({\it ii}) a substructure within the Sagittarius Arm, or ({\it iii}) an inter-arm spur warrants searches for similar structures along the $\ell$--$v$ locus of the Sagittarius Arm, using VLBI, Gaia parallaxes, and dust extinction distances. This new structure in Sagittarius provides an excellent laboratory for examining star formation on scales large enough to be compared to extragalactic observations, but with the ability to resolve the mass function, spatial distribution, and kinematics of the individual sources.

\begin{appendix}

\section{History of the Sagittarius Arm}\label{sec:history}

The characterization of the high-pitch-angle,  kiloparsec-long, star-forming structure described here is unremarkable in the context of spiral galaxies; numerous galaxies show similar structures, some related to an overall spiral pattern and some whose origin is not so clear.  But as we discuss here, this structure is indeed unprecedented in the context of the generally adopted model of the Milky Way spiral structure. In this appendix we outline the development of the Sagittarius Arm in the astronomical literature. This is by no means a complete accounting of its properties, but provides insight into the development of currently used models. 

Our definition of the Sagittarius Arm had its beginnings with the first generally accepted map of spiral structure, presented by W.\ W.\ Morgan at the 86th meeting of the American Astronomical Society in December 1951.  At this meeting, Morgan presented slides of a physical model he had built at Yerkes Observatory: a board into which he had pounded 25 nails, each topped with white balls to indicate his measured distance to an OB association. The focus of this presentation was a band of star formation near the Sun, later called the Orion Arm, or Local Arm, and regions of star formation in what was later dubbed the Perseus Arm.  This board had a single point inward from the Sun's position: the start of the Sagittarius Arm \citep{Gingerich1985}.

Although the data from this original presentation were never published, an expanded version of this work was published by \citet{Morgan1953}. By this point, the number of inner Galaxy OB associations had grown to seven, with measurements for eight additional single stars.  Details on the full stellar sample were published in a companion paper \citep{Morgan1955}. We recalculated the distances to these stars using \textit{Gaia} EDR3 and calculated the average distance to revise the distance of each association (Fig.~\ref{fig:GLIMPSE}). Modern parallaxes indicate that four stars are interlopers; these were removed from the averaging.  We find that six of the original associations mapped in \citet{Morgan1953} belong to the structure that we are characterizing. Surprisingly, the high pitch angle is clear in this original 1953 work but went unremarked. We searched the literature for any discussion of the depth of this structure and only found two manuscripts that clearly discuss it, both of which attribute it to a variable thickness for the arm \citep{Avedisova1989,Gerasimenko1993}. Evidence for a high pitch angle structure in this direction was also provided in the review talk by D. Elmegreen at {\it IAU Symposium 106: The Milky Way Galaxy} and was remarked upon at that meeting \citep{Elmegreen1985}.

Given the uncertainties in distances, the patchiness of extinction in the inner Galaxy, and the intense focus on finding a spiral structure similar to other spirals -- with a lower pitch angle -- it is unsurprising that the alignment of these objects did not draw attention. The advent of 21 cm astronomy, which began with the detection of the hyperfine transition of HI in March, May, and July 1951 \citep{Ewen1951,Muller1951,Pawsey1951}, focused the community on the larger-scale structure of spiral arms, beginning with the pioneering paper of \citet{vanDeHulst1954}. Although this paper focused on the distribution of neutral hydrogen outside the solar circle, the results were compared with those of Morgan, Whitford, and Code. Van de Hulst, Muller, and Oort -- in consultation with Morgan -- proposed the names {\it Perseus Arm}, {\it Orion Arm}, and {\it Sagittarius Arm} for the concentrations of star formation mapped by Morgan. 

The first explicit linkage between the Sagittarius Arm OB associations and the HI data came in \citet{Kwee1954}. This paper focused on the HI rotation curve, measuring the maximum radial velocity along the line of sight. They found that a plot of this terminal velocity as a function of galactocentric radius showed three prominent dips, one of which they associated with the same range of galactocentric radius as the Sagittarius OB associations. They speculated that these two regions of the Galaxy were connected, as illustrated in Fig. 8 of \citet{Kwee1954}. The idea that the $l=49^{\circ}$ direction marked the point where the arm went into tangency was reinforced by the discovery of W51 by \citet{Westerhout1958}.  This optically obscured but bright radio continuum source has since been frequently attributed to arising along the tangency of the Sagittarius Arm. Modern measurements of VLBI parallaxes toward seven objects in this direction confirm that the star formation is spread out over 2 kpc, as one might expect for a spiral arm tangency \citep{Reid2019}. A study of 21 cm emission by \citet{Schmidt1957} claimed evidence for HI gas in the inner Galaxy that was assumed to be associated with the Sagittarius Arm, although the association with the stellar associations was not documented. 

Current spiral structure models principally derive from efforts to identify continuous bands of high intensity emission as a function of galactic longitude and radial velocity: the $\ell$--$v$ diagram. The expectation was to find loops in this diagram, where the turnaround occurs as the arm transitions from near kinematic distances to far kinematic distances as it reaches tangency.  These efforts proceeded using 21 cm emission, and then emission from CO. The first  $\ell$--$v$ track tracing the Sagittarius Arm -- to our knowledge -- was in Fig. 5 of the 21 cm study of \citet{Burton1966}, which identified a possible loop over the range $\ell=43^{\circ}$ to $53^{\circ}$.  In this paper, Burton noted the presence of emission at higher velocities than those of the Sagittarius Arm loop, which led to subsequent investigations of the role of streaming motions due to a spiral density wave -- as opposed to overdensities -- in creating features in the $\ell$--$v$ diagram \citep{Burton1971,Burton1972,Burton1974}.

The $\ell$--$v$ track identified by Burton was extrapolated further into the inner galaxy with the inclusion of models as well as reference to unpublished data from S.\ C.\ Simonson \citep{Burton1970}. This Sagittarius Arm $\ell$--$v$ track was revived with the CO investigations of \citet{Cohen1980} and extended in the highly influential work of \citet{Dame1986}, which identified 17 large molecular complexes distributed rather uniformly along a 15 kpc stretch. An oft reproduced image from that paper is their Fig. 10, which gives the impression of the Sagittarius Arm having molecular clouds spaced like ``beads on a string'' lying along a logarithmic spiral with pitch angle 5.3$^\circ$. Using VLBI masers parallaxes, this was revised to $6.9\pm1.6$ degrees \citep{Wu2014,Reid2014}.

A particularly problematic aspect of constructing a spiral locus for the Sagittarius Arm has been determining how it extends into the fourth quadrant ($\ell=360-270^{\circ})$. This is challenging because the kinematic method of estimating the distance to sources becomes degenerate toward the Galactic center, disconnecting structures identified in the first and fourth quadrants.  A pivotal moment in this discussion occurred during and after IAU Symposium No.\ 38 in Basel in 1969. At this meeting, two maps constructed from H\,{\sc i} observations, one by \citet{Kerr1969} and one by \citet{Weaver1970}, appeared to have very different characteristics. At a workshop the following year, it was noted that discrepancies in the inner Galaxy were related to spiral arm tangent points and the connections between them \citep{Simonson1970}. The identification of tangency directions has always played an important part in informing models of the spiral structure, starting with the work of \citet{Mills1959} and continuing through to the present day \citep{Hou2015}.

At the 1969 symposium, Kerr chose to connect the Sagittarius Arm tangency at $\ell=49^\circ$ to the $\ell=305^\circ$ tangency (Centaurus--Crux) in the fourth quadrant, while Weaver linked the Sagittarius Arm to the Carina tangency at $\ell=283^\circ$. The latter choice ended up propagating through to most current models, which require a pitch angle of $\sim 12$ degrees. 

The tension introduced by fitting local sections of spiral arms and attempting to fit a larger, suspected log-spiral pattern based on tangency directions can be seen in the most recent paper of \citet{Reid2019}. By introducing a ``kink'' in the Sagittarius-Carina spiral arm at a galactocentric angle 24 degrees clockwise from the Galactic-center-to-Sun direction, they find a pitch angle of only one degree for the Sagittarius Arm in most of the first quadrant and a pitch angle of 17 degrees for its extension into the fourth quadrant (where no VLBI parallax data are available). The deviation of a Sagittarius-Carina arm from an idealized log-spiral had been previously proposed based on measurements of photometric distances to the exciting stars of associated H\,{\sc ii} regions. As an example, the well-known \citet{TaylorCordes93} model of the spatial distribution of free electrons started with the \citet{Georgelin1976} logarithmic-spiral model but incorporated a kink based on the H\,{\sc ii} regions distance measurements of \citet{Downes1980}. However, none of these proposed revisions seems to have gained general usage. For convenience, most models have continued to use the log-spiral approximation. 

\section{Bayesian model of cluster astrometry}\label{sec:model}

To estimate the basic astrometric properties of a YSO group, we used a mixture model -- a probability distribution made by adding several components -- that also takes into account the heteroscedastic measurement uncertainties tabulated by Gaia EDR3. Each component of this model is astrophysically motivated, with one (or multiple) component(s) corresponding the star cluster and another component corresponding to non-clustered contaminants found along the same line of sight. 

Clustered stars must be at the same distance, and Gaia parallax measurement errors are approximately Gaussian \citep{Lindegren2020_as}. Thus, for each member star, we assumed identical mean parallaxes, $\varpi_0$, and standard deviations obtained from the Gaia EDR3 tables. We also expect member stars to share similar proper motions, but, in addition to proper motion measurement error, clusters have an internal velocity dispersion that produce small spreads in proper motions. Thus, for proper motions, we used $t$ distributions rather than Gaussian distributions because the heavy tails provide increased robustness to cluster members with discrepant proper motion. For these $t$ distributions, we treated the degrees of freedom, $\nu$, as a nuisance parameter that we marginalized over; $\nu$ was fixed to be the same for both $\mu_{\ell,^\star}$ and $\mu_b$. 

For star $i$ of a cluster, the probability distribution is
\begin{multline}
    p_\mathrm{clust}(\varpi_{i},\mu_{{\ell^\star},i},\mu_{{b},i}|\varpi_0, \mu_{\ell^\star,0}, \mu_{b,0}) =\\ \phi(\varpi_{i}|\varpi_0,\sigma_{\varpi_{i}}^2)\cdot
    {f}(\mu_{{\ell^\star},i}|\mu_{\ell^\star,0},\sigma_{\mu_{{\ell^\star},0}}^2,\nu_\mu)\cdot
    {f}(\mu_{{b},i}|\mu_{b,0},\sigma_{\mu_{{b},0}}^2,\nu_\mu),
\end{multline}
where $\theta = (\varpi_0, \mu_{\ell^\star,0}, \mu_{b,0})$ are the mean astrometric values for the cluster, $x_i = (\varpi_i, \mu_{\ell^\star,i}, \mu_{b,i})$ are the measured values for the $i$th star, $\sigma_i$ are corresponding uncertainties, $\phi$ denotes a Gaussian distribution, and $f$ denotes a $t$ distribution. In our mixture model, we included either one or two cluster components, which correspond to the cases where there is either a single cluster or two clusters superimposed along the line of sight. 

For contaminants, we expect the distributions of $\varpi$, $\mu_{\ell,^\star}$, and $\mu_b$ to be broader than for cluster members, but these distributions may shift depending on the direction on the sky. Appendix~\ref{sec:non-yso} approximates these distributions, $p_\mathrm{contam}(x_i|\ell,b)$, as a function of Galactic coordinates.

For a YSO group that is assumed to be a single cluster, the likelihood equation is
\begin{equation}
    p(x_i|\theta) = a\,p_\mathrm{clust}(x_i|\theta) + (1-a)\,p_\mathrm{contam}(x_i|\ell,b),
\end{equation}
where $0\leq a \leq 1$ is the mixing parameter indicating the fraction of stars that are {\it bona fide} cluster members rather than non-clustered contaminants. In the case of two clusters, a second cluster component is added to the above equation.

The prior distributions for our Bayesian model are
\begin{align}
\label{eq:priors}
   d_{\odot} &\sim {\rm Uniform}(0,8000),  \\ 
   \mu_{\ell^\star} &\sim {\rm Uniform}(-10,4), \notag \\ 
   \mu_b &\sim {\rm Uniform}(-5,2), \notag \\ 
       a &\sim \mathrm{Beta}(2,0.5),\notag\\
   \nu_\mu &\sim {\rm Gamma}(2,0.2), \notag  
    \end{align}
where $d_{\odot}~[\mathrm{pc}] = 1000/\varpi_0~[\mathrm{mas}]$ is the cluster heliocentric distance.
For the quantities of interest, we used uniform priors within reasonable astrophysical ranges.
For the mixing parameter, we adopted a prior that mildly favors low contamination rates because previous studies of clustered YSOs selected in similar ways have found contamination rates of $\sim$20\% \citep[e.g.,][]{Kuhn2019}.  For $\nu_\mu$, we adopted a prior that permits $t$ distributions with heavy wings. 

For each YSO group, we ran three MCMC chains with JAGS \citep{Plummer2019} via the R2jags package \citep{Su2020}. Convergence was assessed by checking that the \citet{Gelman1992} statistic is $<$1.001. The 95\% credible intervals for each astrometric parameter are reported in Table~\ref{tab:structure}. We also report the number of stars, $N_{mem}$, that have a $>$50\% probability of belonging to the cluster component of the model.

Among the YSO groups investigated in this letter, there are two cases where including a second cluster component significantly improved the model. These are G29.9+2.2 (not part of the structure), where two clusters at different distances are aligned along the line of sight, and G14.1-0.5 ($=$~M17 SWex), where stars in different parts of the molecular cloud have different proper motions. In both cases the Bayes factor strongly favored the more complex model  ($BF > 10^{10}$).  

\section{Astrometric properties of nonmembers}\label{sec:non-yso}

Young stellar object groups identified by HDBSCAN $(\ell,b)$ clustering may include nonmembers. Here, we justify our assumption from Appendix~\ref{sec:model} that the astrometric properties of these objects can be modeled by $t$ distributions. If these interlopers are mostly non-YSO contaminants, we might expect them to have properties similar to the $\sim$200,000 red infrared sources classified as non-YSOs in \citetalias{spicyI}. Thus, we used this sample to model the distributions.  

\begin{figure}
        \centering
        \includegraphics[width=0.47\textwidth]{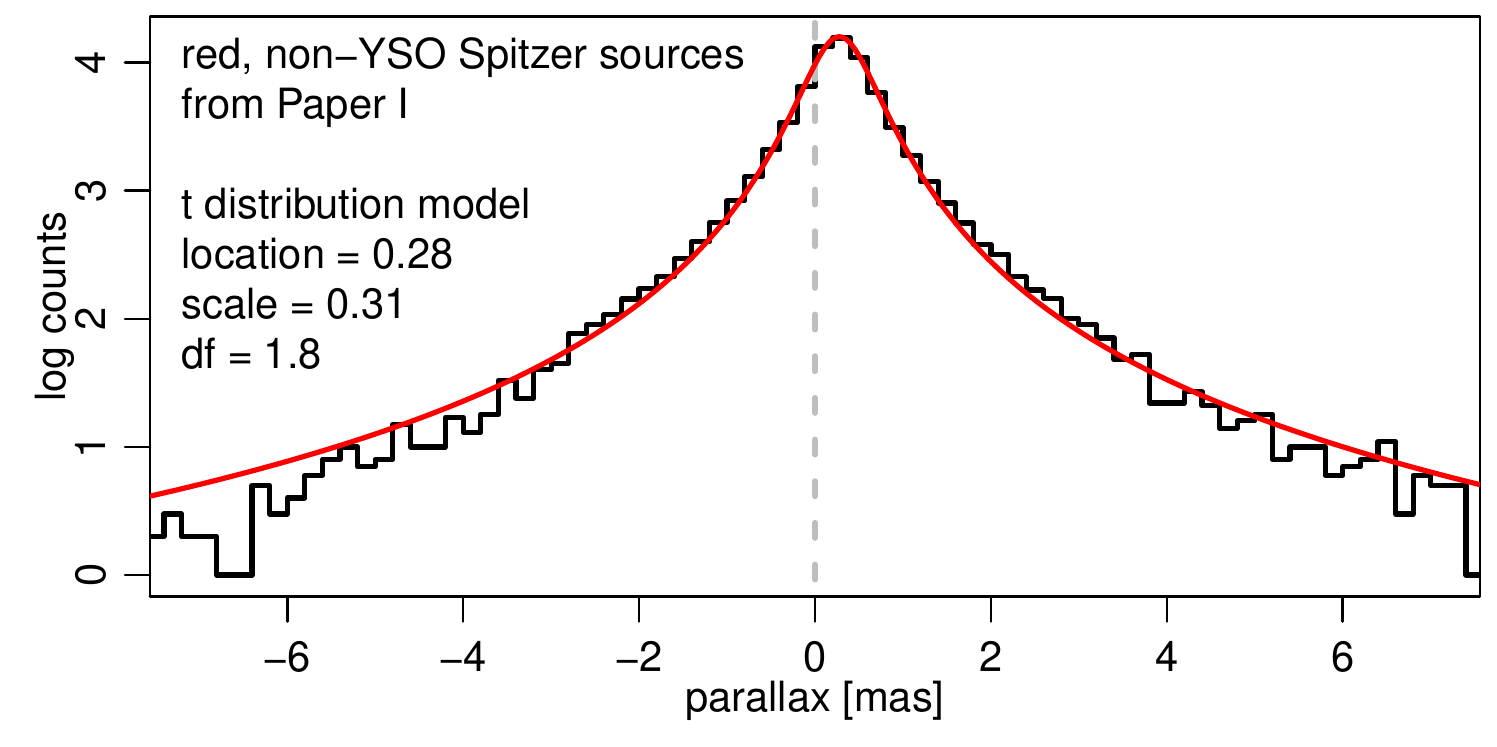}
    \caption{Histogram of Gaia EDR3 parallaxes for red mid-infrared sources from \citet{spicyI} classified as non-YSOs. A $t$ distribution (red curve) is able to account for the heavy tails.}
  \label{fig:t_nonmem}
\end{figure}

Figure~\ref{fig:t_nonmem} shows the parallax distribution of the full non-YSO sample. This distribution, produced by the combination of variations in distance and heteroscedastic measurement errors, is clearly non-Gaussian, but the heavy tails appear to be well described by the overplotted $t$ distribution. Similar tails are found for $\mu_{\ell^\star}$ and $\mu_b$, but the distribution centers are not as accurately modeled owing to proper motion variations with Galactic $\ell$.

To more accurately approximate the $\varpi$, $\mu_{\ell^\star}$, and $\mu_b$ distributions as functions of $(\ell, b),$ we divided the survey area into boxes with $\Delta\ell$=10$^\circ$ and $\Delta b$=40$^\prime$ and fit the non-YSOs in each box with $t$ distributions using the {\tt fitdistr} function from the {\it R} library {\it MASS} \citep{VenablesRipley2002}. We then used the local polynomial regression algorithm {\tt loess} \citep{Cleveland1992} to estimate $t$-distribution parameters as smooth functions of $(\ell, b)$. Here we adopted the {\tt loess} implementation in base {\it R} v3.6.0 \citep{RCore} with {\tt span~$=$~0.2} and {\tt degree~$=$~2}. Examination of probability--probability plots (not shown) indicates that this provides excellent approximations for non-YSOs distributions.

\section{Properties of objects in the linear structure}\label{sec:table}

The YSO groups comprising the structure are listed, along with their properties, in Tables~\ref{tab:structure} and \ref{tab:structure2}. The quantities in Table~\ref{tab:structure} include: (i) the number of stars with Gaia data, $N_\mathrm{Gaia}$, (ii) the number of those stars classified as probable members, $N_\mathrm{mem}$, (iii) the centers of the groups' $(l_0, b_0)$ in Galactic coordinates, (iv) the mean parallax, $\varpi_0$, heliocentric distance, $d_\odot$, and mean proper motions, $(\mu_{\ell^\star,0}, \mu_{b,0})$, in Galactic coordinates, and (v) the groups' radial velocities, $v_\mathrm{lsr}$, in the ``standard'' local standard of rest.\footnote{By historical convention, $v_\mathrm{lsr}$ are reported in a reference frame that takes the solar peculiar motion to be 20 km~s$^{-1}$ toward 20$^\mathrm{h}$ 30$^\circ$ (B1900).}

To facilitate comparison with the positions and kinematics of the masers of \citet{Reid2019}, we adopted the same Galactic parameters here -- $R_0=8.15$~kpc, $Z_0 = +5.5$~pc and a solar motion of $UVW=(10.6,10.7,7.6)$~km~s$^{-1}$ -- and compared the space motions of the YSO groups to their derived rotation curve with circular speed, $\Theta = 236$~km~s$^{-1}$, at the position of the Sun.   Other authors prefer different parameters  \citep[e.g.,][]{GravityCollaboration2019,Eilers2019}, which would produce a small but systematic change in our derived values.

The quantities in Table~\ref{tab:structure2} include: (i) the heliocentric Cartesian coordinates defined as $x =  d_\odot \cos(\ell_0)\cos(b_0)$, $y=d_\odot \sin(\ell_0)\cos(b_0)$, and $z = d_\odot \sin(b_0)$, (ii) the heliocentric Cartesian velocities $(v_x,v_y, v_z)$, (iii) the galactocentric Cartesian positions and velocities $(X, Y, Z, V_X, V_Y, V_Z)$, and (iv) the galactocentric positions and velocities in cylindrical coordinates, where $\varphi$ is the azimuthal angle, $V_R$ is the radial velocity with respect to the Galactic center, and $|V_\varphi|$ is the azimuthal velocity. Uncertainties take into account the astrometric uncertainties and uncertainties on $v_\mathrm{lsr}$ described in Appendix~\ref{sec:CO}.

\begin{sidewaystable*}[b]
\caption{YSO groups in the Sagittarius Arm structure.}\label{tab:structure}
\centering
\begin{tabular}{lrrrrrrrrl} 
\hline\hline       
Group & 
\multicolumn{1}{c}{$N_\mathrm{Gaia}$} & 
\multicolumn{1}{c}{$N_\mathrm{Mem}$} & 
\multicolumn{1}{c}{$\ell_0$} & 
\multicolumn{1}{c}{$b_0$} & 
\multicolumn{1}{c}{$\varpi_0$} & 
\multicolumn{1}{c}{$d_\odot$} & 
\multicolumn{1}{c}{$\mu_{\ell^\star,0}$} &
\multicolumn{1}{c}{$\mu_{b,0}$} & 
\multicolumn{1}{c}{$v_\mathrm{lsr}$} \\
&&&
\multicolumn{1}{c}{($^\circ$)}&
\multicolumn{1}{c}{($^\circ$)}&
\multicolumn{1}{c}{(mas)}&
\multicolumn{1}{c}{(kpc)}&
\multicolumn{1}{c}{(mas~yr$^{-1}$)}&
\multicolumn{1}{c}{(mas~yr$^{-1}$)}&
\multicolumn{1}{c}{(km~s$^{-1}$)}\\
\hline
G4.0-1.0 & 42 & 30 & $ 4.07$ & $-1.00$ & $0.856$ ($0.803$, $0.909$) & $1.17$ ($1.10$, $1.24$) & $-0.725$ ($-0.837$, $-0.580$) & $-2.928$ ($-3.001$, $-2.801$) & 11.7\\
G5.9-0.9 & 31 & 22 & $ 5.93$ & $-0.98$ & $0.774$ ($0.734$, $0.811$) & $1.29$ ($1.23$, $1.36$) & $-0.832$ ($-0.919$, $-0.758$) & $-2.490$ ($-2.576$, $-2.416$) & 13.0\\
G6.0-1.3 (M8) & 246 & 225 & $ 6.06$ & $-1.35$ & $0.821$ ($0.810$, $0.833$) & $1.22$ ($1.20$, $1.23$) & $-1.181$ ($-1.200$, $-1.159$) & $-2.193$ ($-2.244$, $-2.153$) & 11.7\\
G6.6-1.5 & 21 & 20 & $ 6.67$ & $-1.53$ & $0.873$ ($0.811$, $0.934$) & $1.15$ ($1.07$, $1.23$) & $-0.411$ ($-0.481$, $-0.333$) & $-2.006$ ($-2.068$, $-1.937$) & 13.0\\
G6.8-2.3 (NGC 6559) & 37 & 30 & $ 6.89$ & $-2.35$ & $0.838$ ($0.787$, $0.892$) & $1.19$ ($1.12$, $1.27$) & $-0.628$ ($-0.705$, $-0.546$) & $-2.485$ ($-2.577$, $-2.392$) & 11.7\\
G6.9-0.2 (M20) & 43 & 37 & $ 6.97$ & $-0.25$ & $0.847$ ($0.811$, $0.885$) & $1.18$ ($1.13$, $1.23$) & $-1.001$ ($-1.136$, $-0.796$) & $-1.568$ ($-1.735$, $-1.434$) & 8.0, 18.5\\
G7.3-1.9 (IC 1274) & 73 & 64 & $ 7.32$ & $-2.00$ & $0.844$ ($0.816$, $0.872$) & $1.18$ ($1.15$, $1.23$) & $-0.388$ ($-0.433$, $-0.344$) & $-2.411$ ($-2.475$, $-2.342$) & 9.1\\
G8.4-0.2 & 9 & 8 & $ 8.44$ & $-0.28$ & $0.735$ ($0.157$, $1.101$) & $1.36$ ($0.91$, $6.35$) & $-1.293$ ($-9.318$, $-1.135$) & $-0.332$ ($-0.729$, $ +1.892$) & 16.5, 36.8\\
G8.5-0.9 (Majaess 190) & 24 & 21 & $ 8.52$ & $-1.00$ & $0.789$ ($0.747$, $0.832$) & $1.27$ ($1.20$, $1.34$) & $-0.629$ ($-0.703$, $-0.399$) & $-1.927$ ($-2.164$, $-1.746$) & 15.6\\
G8.5-0.6 & 22 & 19 & $ 8.52$ & $-0.63$ & $0.716$ ($0.652$, $0.789$) & $1.40$ ($1.27$, $1.53$) & $-0.799$ ($-0.879$, $-0.706$) & $-2.072$ ($-2.153$, $-1.969$) & 16.9\\
G10.3-0.8 & 49 & 42 & $10.33$ & $-0.83$ & $0.684$ ($0.633$, $0.736$) & $1.46$ ($1.36$, $1.58$) & $-1.646$ ($-1.725$, $-1.582$) & $-1.823$ ($-1.883$, $-1.762$) & 18.2, 28.6\\
G10.3+2.2 & 26 & 19 & $10.40$ & $ 2.25$ & $0.736$ ($0.616$, $0.826$) & $1.36$ ($1.21$, $1.62$) & $-1.559$ ($-1.724$, $-1.273$) & $-1.557$ ($-1.683$, $-1.461$) & 6.5, 23.4\\
G10.8-2.4 (LDN 291) & 131 & 103 & $10.87$ & $-2.43$ & $0.788$ ($0.769$, $0.806$) & $1.27$ ($1.24$, $1.30$) & $-0.684$ ($-0.761$, $-0.620$) & $-1.635$ ($-1.692$, $-1.574$) & 11.7\\
G12.3+0.4 & 14 & 13 & $12.40$ & $ 0.50$ & $0.662$ ($0.513$, $0.809$) & $1.51$ ($1.24$, $1.95$) & $-1.100$ ($-1.349$, $-0.758$) & $-1.923$ ($-2.160$, $-1.647$) & 18.0, 31.0\\
G12.7+0.4 (LBN 49) & 21 & 17 & $12.78$ & $ 0.42$ & $0.544$ ($0.437$, $0.649$) & $1.84$ ($1.54$, $2.29$) & $-0.786$ ($-1.106$, $-0.624$) & $-1.329$ ($-1.597$, $-1.036$) & 19.0\\
G13.3+0.1 & 13 & 11 & $13.34$ & $ 0.19$ & $0.492$ ($0.316$, $0.668$) & $2.03$ ($1.50$, $3.17$) & $-1.435$ ($-1.616$, $-1.280$) & $-1.676$ ($-1.838$, $-1.520$) & 18.0\\
G14.1-0.5 A (M17 SWex)& 119 & 48 & $14.19$ & $-0.53$ & $0.629$ ($0.572$, $0.676$) & $1.59$ ($1.48$, $1.75$) & $-1.704$ ($-1.808$, $-1.567$) & $-0.867$ ($-0.928$, $-0.762$) & 19.9\\
G14.1-0.5 B (M17 SWex)& 119 & 40 & $14.19$ & $-0.53$ & $0.621$ ($0.584$, $0.660$) & $1.61$ ($1.52$, $1.71$) & $-1.230$ ($-1.292$, $-1.138$) & $-1.600$ ($-1.649$, $-1.553$) & 19.9\\
G14.5-0.1 & 37 & 10 & $14.56$ & $-0.11$ & $0.538$ ($0.471$, $0.603$) & $1.86$ ($1.66$, $2.12$) & $-1.024$ ($-1.107$, $-0.951$) & $-1.409$ ($-1.519$, $-0.958$) & 25.5, 39.8\\
G15.1-0.5 (M17) & 15 & 11 & $15.11$ & $-0.59$ & $0.649$ ($0.587$, $0.744$) & $1.54$ ($1.34$, $1.70$) & $-1.479$ ($-1.580$, $-1.413$) & $-1.037$ ($-1.410$, $-0.864$) & 19.3\\
G15.6-0.3 & 25 & 11 & $15.61$ & $-0.34$ & $0.640$ ($0.458$, $0.707$) & $1.56$ ($1.42$, $2.19$) & $-1.550$ ($-1.617$, $-1.448$) & $-1.052$ ($-1.167$, $-0.828$) & 16.0, 20.5,\\
&&&&&&&&&40.0, 58.0\\
G16.6-0.1 & 78 & 56 & $16.63$ & $-0.19$ & $0.533$ ($0.432$, $0.626$) & $1.88$ ($1.60$, $2.31$) & $-1.557$ ($-1.799$, $-1.365$) & $-0.583$ ($-0.730$, $-0.475$) & 24.3, 47.5,\\
&&&&&&&&&56.3\\
G16.9+0.8 (M16) & 134 & 121 & $16.97$ & $ 0.81$ & $0.601$ ($0.577$, $0.625$) & $1.66$ ($1.60$, $1.73$) & $-1.315$ ($-1.359$, $-1.265$) & $-0.963$ ($-1.005$, $-0.921$) & 20.6\\
G18.0+0.2 & 16 & 8 & $18.09$ & $ 0.22$ & $0.489$ ($0.261$, $0.654$) & $2.04$ ($1.53$, $3.84$) & $-0.543$ ($-1.675$, $-0.352$) & $-0.540$ ($-0.782$, $-0.363$) & 22.5\\
G18.4+1.9 (Sh 2-54) & 463 & 395 & $18.45$ & $ 1.91$ & $0.512$ ($0.499$, $0.525$) & $1.95$ ($1.90$, $2.01$) & $-2.127$ ($-2.157$, $-2.101$) & $-0.642$ ($-0.675$, $-0.608$) & 27.3\\
\hline
\end{tabular}
\tablefoot{
Ranges for values indicate the 95\% credible intervals from Bayesian analysis. All plausible $v_\mathrm{lsr}$ solutions are listed.}
\end{sidewaystable*}

\setlength{\tabcolsep}{2pt}
\begin{sidewaystable*}
\caption{Kinematics of YSO groups.}\label{tab:structure2}
\centering\tiny
\begin{tabular}{lrrrrrrrrrrrrrrrrrr} 
\hline\hline       
& \multicolumn{6}{c}{Heliocentric} && \multicolumn{11}{c}{galactocentric}\\ \cline{2-7}  
& &&&&&&&\multicolumn{6}{c}{Cartesian}&& \multicolumn{4}{c}{Cylindrical}\\ \cline{9-14} \cline{16-19}
Group & 
\multicolumn{1}{c}{x} & 
\multicolumn{1}{c}{y} & 
\multicolumn{1}{c}{z} & 
\multicolumn{1}{c}{$v_x$} & 
\multicolumn{1}{c}{$v_y$} & 
\multicolumn{1}{c}{$v_z$} && 
\multicolumn{1}{c}{X} & 
\multicolumn{1}{c}{Y} & 
\multicolumn{1}{c}{Z} & 
\multicolumn{1}{c}{$V_X$} & 
\multicolumn{1}{c}{$V_Y$} & 
\multicolumn{1}{c}{$V_Z$} && 
\multicolumn{1}{c}{R} & 
\multicolumn{1}{c}{$\varphi$} & 
\multicolumn{1}{c}{$V_R$} & 
\multicolumn{1}{c}{$|V_\varphi|$} \\
&
\multicolumn{1}{c}{(pc)}&
\multicolumn{1}{c}{(pc)}&
\multicolumn{1}{c}{(pc)}&
\multicolumn{1}{c}{(km~s$^{-1}$)}&
\multicolumn{1}{c}{(km~s$^{-1}$)}&
\multicolumn{1}{c}{(km~s$^{-1}$)}&&
\multicolumn{1}{c}{(pc)}&
\multicolumn{1}{c}{(pc)}&
\multicolumn{1}{c}{(pc)}&
\multicolumn{1}{c}{(km~s$^{-1}$)}&
\multicolumn{1}{c}{(km~s$^{-1}$)}&
\multicolumn{1}{c}{(km~s$^{-1}$)}&&
\multicolumn{1}{c}{(pc)}&
\multicolumn{1}{c}{($^\circ$)}&
\multicolumn{1}{c}{(km~s$^{-1}$)}&
\multicolumn{1}{c}{(km~s$^{-1}$)}\\
\hline
G4.0-1.0 & $ 1165\pm   35$ & $   83\pm    2$ & $  -20\pm    1$ & $   0.5\pm   5.2$ & $  -4.0\pm   0.5$ & $ -16.2\pm   0.6$ &  & $-6985\pm   35$ & $   83\pm    2$ & $  -15\pm    1$ & $  11.1\pm   5.2$ & $ 242.7\pm   0.5$ & $  -8.6\pm   0.6$ &  & $ 6986\pm   35$ & 179.3 & $  -8.1\pm   5.2$ & $ 242.8\pm   0.5$ \\
G5.9-0.9 & $ 1284\pm   31$ & $  133\pm    3$ & $  -22\pm    1$ & $   1.6\pm   5.0$ & $  -5.0\pm   0.6$ & $ -15.3\pm   0.5$ &  & $-6866\pm   31$ & $  133\pm    3$ & $  -17\pm    1$ & $  12.2\pm   5.0$ & $ 241.7\pm   0.6$ & $  -7.7\pm   0.5$ &  & $ 6867\pm   31$ & 178.9 & $  -7.3\pm   4.9$ & $ 241.9\pm   0.7$ \\
G6.0-1.3 & $ 1210\pm    8$ & $  128\pm    1$ & $  -29\pm    1$ & $   0.5\pm   4.8$ & $  -6.8\pm   0.5$ & $ -12.7\pm   0.2$ &  & $-6940\pm    8$ & $  128\pm    1$ & $  -23\pm    1$ & $  11.1\pm   4.8$ & $ 239.9\pm   0.5$ & $  -5.1\pm   0.2$ &  & $ 6941\pm    8$ & 179.0 & $  -6.5\pm   4.8$ & $ 240.1\pm   0.6$ \\
G6.6-1.5 & $ 1137\pm   40$ & $  133\pm    5$ & $  -30\pm    1$ & $   1.2\pm   4.7$ & $  -2.1\pm   0.6$ & $ -10.9\pm   0.4$ &  & $-7013\pm   40$ & $  133\pm    5$ & $  -25\pm    1$ & $  11.8\pm   4.7$ & $ 244.6\pm   0.6$ & $  -3.3\pm   0.4$ &  & $ 7014\pm   40$ & 178.9 & $  -7.0\pm   4.7$ & $ 244.8\pm   0.7$ \\
G6.8-2.3 & $ 1183\pm   38$ & $  143\pm    5$ & $  -49\pm    2$ & $  -0.2\pm   5.1$ & $  -3.6\pm   0.6$ & $ -14.0\pm   0.6$ &  & $-6967\pm   38$ & $  143\pm    5$ & $  -43\pm    2$ & $  10.4\pm   5.1$ & $ 243.1\pm   0.6$ & $  -6.4\pm   0.6$ &  & $ 6968\pm   38$ & 178.9 & $  -5.3\pm   5.1$ & $ 243.3\pm   0.7$ \\
G6.9-0.2 & $ 1171\pm   25$ & $  143\pm    3$ & $   -5\pm    1$ & $  -3.4\pm   5.8$ & $  -6.1\pm   0.8$ & $  -8.8\pm   0.5$ &  & $-6979\pm   25$ & $  143\pm    3$ & $    0\pm    1$ & $   7.2\pm   5.8$ & $ 240.6\pm   0.8$ & $  -1.2\pm   0.5$ &  & $ 6980\pm   25$ & 178.8 & $  -2.1\pm   5.8$ & $ 240.7\pm   0.9$ \\
G7.3-1.9 & $ 1174\pm   19$ & $  151\pm    2$ & $  -41\pm    1$ & $  -3.0\pm   4.9$ & $  -2.6\pm   0.7$ & $ -13.4\pm   0.3$ &  & $-6976\pm   19$ & $  151\pm    2$ & $  -36\pm    1$ & $   7.6\pm   4.9$ & $ 244.1\pm   0.7$ & $  -5.8\pm   0.3$ &  & $ 6978\pm   19$ & 178.8 & $  -2.2\pm   4.8$ & $ 244.2\pm   0.8$ \\
G8.4-0.2 & $ 1346\pm  382$ & $  200\pm   57$ & $   -7\pm    2$ & $   5.3\pm   8.9$ & $  -7.6\pm  13.0$ & $  -2.2\pm   4.4$ &  & $-6804\pm  382$ & $  200\pm   57$ & $   -1\pm    2$ & $  15.9\pm   8.9$ & $ 239.1\pm  13.0$ & $   5.4\pm   4.4$ &  & $ 6807\pm  379$ & 178.4 & $  -8.7\pm   8.8$ & $ 239.4\pm  13.0$ \\
G8.5-0.9 & $ 1254\pm   34$ & $  188\pm    5$ & $  -22\pm    1$ & $   3.6\pm   5.0$ & $  -3.3\pm   0.9$ & $ -11.6\pm   0.7$ &  & $-6896\pm   34$ & $  188\pm    5$ & $  -17\pm    1$ & $  14.2\pm   5.0$ & $ 243.4\pm   0.9$ & $  -4.0\pm   0.7$ &  & $ 6899\pm   34$ & 178.5 & $  -7.4\pm   5.0$ & $ 243.7\pm   1.0$ \\
G8.5-0.6 & $ 1381\pm   63$ & $  207\pm   10$ & $  -15\pm    1$ & $   5.1\pm   4.9$ & $  -4.6\pm   0.8$ & $ -13.8\pm   0.7$ &  & $-6769\pm   63$ & $  207\pm   10$ & $  -10\pm    1$ & $  15.7\pm   4.9$ & $ 242.1\pm   0.8$ & $  -6.2\pm   0.7$ &  & $ 6773\pm   63$ & 178.3 & $  -8.1\pm   4.8$ & $ 242.5\pm   0.9$ \\
G10.3-0.8 & $ 1438\pm   57$ & $  262\pm   10$ & $  -21\pm    1$ & $   7.2\pm   5.7$ & $ -10.3\pm   1.1$ & $ -12.7\pm   0.5$ &  & $-6712\pm   57$ & $  262\pm   10$ & $  -16\pm    1$ & $  17.8\pm   5.7$ & $ 236.4\pm   1.1$ & $  -5.1\pm   0.5$ &  & $ 6717\pm   57$ & 177.8 & $  -8.4\pm   5.6$ & $ 236.9\pm   1.3$ \\
G10.3+2.2 & $ 1335\pm   90$ & $  245\pm   17$ & $   53\pm    4$ & $  -4.4\pm   7.4$ & $ -11.0\pm   1.7$ & $ -10.3\pm   0.8$ &  & $-6815\pm   90$ & $  245\pm   17$ & $   59\pm    4$ & $   6.2\pm   7.4$ & $ 235.7\pm   1.7$ & $  -2.7\pm   0.8$ &  & $ 6820\pm   90$ & 178.0 & $   2.5\pm   7.4$ & $ 235.8\pm   1.8$ \\
G10.8-2.4 & $ 1245\pm   15$ & $  239\pm    3$ & $  -54\pm    1$ & $  -0.6\pm   4.8$ & $  -4.3\pm   0.9$ & $  -9.8\pm   0.3$ &  & $-6905\pm   15$ & $  239\pm    3$ & $  -48\pm    1$ & $  10.0\pm   4.8$ & $ 242.4\pm   0.9$ & $  -2.2\pm   0.3$ &  & $ 6909\pm   14$ & 178.1 & $  -1.4\pm   4.8$ & $ 242.6\pm   1.1$ \\
G12.3+0.4 & $ 1475\pm  170$ & $  324\pm   37$ & $   13\pm    2$ & $   6.3\pm   5.9$ & $  -6.7\pm   2.0$ & $ -13.7\pm   1.8$ &  & $-6675\pm  170$ & $  324\pm   37$ & $   19\pm    2$ & $  16.9\pm   5.9$ & $ 240.0\pm   2.0$ & $  -6.1\pm   1.8$ &  & $ 6683\pm  168$ & 177.3 & $  -5.0\pm   6.0$ & $ 240.6\pm   2.2$ \\
G12.7+0.4 & $ 1791\pm  181$ & $  406\pm   41$ & $   13\pm    1$ & $   7.0\pm   5.0$ & $  -5.4\pm   1.7$ & $ -11.5\pm   1.7$ &  & $-6359\pm  181$ & $  406\pm   41$ & $   19\pm    1$ & $  17.6\pm   5.0$ & $ 241.3\pm   1.7$ & $  -3.9\pm   1.7$ &  & $ 6372\pm  178$ & 176.4 & $  -1.9\pm   5.3$ & $ 241.9\pm   2.0$ \\
G13.3+0.1 & $ 1976\pm  359$ & $  468\pm   85$ & $    7\pm    1$ & $   7.5\pm   5.1$ & $ -12.4\pm   2.7$ & $ -16.1\pm   3.0$ &  & $-6174\pm  359$ & $  468\pm   85$ & $   12\pm    1$ & $  18.1\pm   5.1$ & $ 234.3\pm   2.7$ & $  -8.5\pm   3.0$ &  & $ 6192\pm  352$ & 175.8 & $  -0.1\pm   5.8$ & $ 235.0\pm   2.9$ \\
G14.1-0.5 A & $ 1541\pm   63$ & $  389\pm   16$ & $  -15\pm    1$ & $   9.1\pm   5.3$ & $ -10.9\pm   1.5$ & $  -6.6\pm   0.5$ &  & $-6609\pm   63$ & $  389\pm   16$ & $   -9\pm    1$ & $  19.7\pm   5.3$ & $ 235.8\pm   1.5$ & $   1.0\pm   0.5$ &  & $ 6620\pm   62$ & 176.7 & $  -5.6\pm   5.1$ & $ 236.5\pm   1.8$ \\
G14.1-0.5 B & $ 1560\pm   51$ & $  394\pm   13$ & $  -15\pm    1$ & $   8.2\pm   4.5$ & $  -7.6\pm   1.3$ & $ -12.3\pm   0.4$ &  & $-6590\pm   51$ & $  394\pm   13$ & $   -9\pm    1$ & $  18.8\pm   4.5$ & $ 239.1\pm   1.3$ & $  -4.7\pm   0.4$ &  & $ 6601\pm   50$ & 176.7 & $  -4.3\pm   4.5$ & $ 239.8\pm   1.5$ \\
G14.5-0.1 & $ 1799\pm  115$ & $  467\pm   30$ & $   -4\pm    1$ & $  13.6\pm   6.1$ & $  -5.8\pm   1.6$ & $ -12.4\pm   1.5$ &  & $-6351\pm  115$ & $  467\pm   30$ & $    2\pm    1$ & $  24.2\pm   6.1$ & $ 240.9\pm   1.6$ & $  -4.8\pm   1.5$ &  & $ 6368\pm  112$ & 175.9 & $  -6.1\pm   6.1$ & $ 242.0\pm   2.0$ \\
G15.1-0.5 & $ 1488\pm   85$ & $  402\pm   23$ & $  -16\pm    1$ & $   8.0\pm   4.7$ & $  -9.0\pm   1.4$ & $  -7.6\pm   1.1$ &  & $-6662\pm   85$ & $  402\pm   23$ & $  -10\pm    1$ & $  18.6\pm   4.7$ & $ 237.7\pm   1.4$ & $  -0.0\pm   1.1$ &  & $ 6675\pm   84$ & 176.6 & $  -4.0\pm   4.8$ & $ 238.4\pm   1.7$ \\
G15.6-0.3 & $ 1506\pm  144$ & $  421\pm   40$ & $   -9\pm    1$ & $   5.0\pm  15.4$ & $ -10.5\pm   4.3$ & $  -7.8\pm   1.0$ &  & $-6644\pm  144$ & $  421\pm   40$ & $   -4\pm    1$ & $  15.6\pm  15.4$ & $ 236.2\pm   4.3$ & $  -0.2\pm   1.0$ &  & $ 6658\pm  141$ & 176.5 & $  -0.3\pm  15.1$ & $ 236.7\pm   5.3$ \\
G16.6-0.1 & $ 1799\pm  176$ & $  537\pm   53$ & $   -6\pm    1$ & $  13.6\pm  11.9$ & $ -10.4\pm   3.7$ & $  -5.2\pm   0.7$ &  & $-6351\pm  176$ & $  537\pm   53$ & $   -1\pm    1$ & $  24.2\pm  11.9$ & $ 236.3\pm   3.7$ & $   2.4\pm   0.7$ &  & $ 6374\pm  171$ & 175.3 & $  -3.8\pm  11.2$ & $ 237.5\pm   4.7$ \\
G16.9+0.8 & $ 1592\pm   32$ & $  486\pm   10$ & $   23\pm    1$ & $   9.0\pm   4.9$ & $  -8.1\pm   1.5$ & $  -7.5\pm   0.2$ &  & $-6558\pm   32$ & $  486\pm   10$ & $   29\pm    1$ & $  19.6\pm   4.9$ & $ 238.6\pm   1.5$ & $   0.1\pm   0.2$ &  & $ 6576\pm   31$ & 175.9 & $  -1.6\pm   4.7$ & $ 239.4\pm   1.9$ \\
G18.0+0.2 & $ 1944\pm  398$ & $  635\pm  130$ & $    8\pm    2$ & $   9.2\pm   4.5$ & $  -2.5\pm   3.5$ & $  -5.2\pm   1.6$ &  & $-6206\pm  398$ & $  635\pm  130$ & $   13\pm    2$ & $  19.8\pm   4.5$ & $ 244.2\pm   3.5$ & $   2.4\pm   1.6$ &  & $ 6239\pm  381$ & 174.3 & $   5.5\pm   8.1$ & $ 244.9\pm   3.7$ \\
G18.4+1.9 & $ 1852\pm   25$ & $  618\pm    8$ & $   65\pm    1$ & $  18.2\pm   4.5$ & $ -14.7\pm   1.6$ & $  -5.5\pm   0.2$ &  & $-6298\pm   25$ & $  618\pm    8$ & $   71\pm    1$ & $  28.8\pm   4.5$ & $ 232.0\pm   1.6$ & $   2.1\pm   0.2$ &  & $ 6328\pm   24$ & 174.5 & $  -5.7\pm   4.4$ & $ 233.7\pm   2.0$ \\
\hline
\end{tabular}
\tablefoot{The entries in this table are derived from the values in Table~\ref{tab:structure}, with conversion to galactocentric coordinates using the Galactic model parameters from \citet{Reid2019}. Uncertainties correspond to one standard deviation and do not include systematic errors in the coordinate system.}
\end{sidewaystable*}

\section{Comparison with the dust extinction}\label{sec:av}

%@arxiver{fE1.png,f3b.pdf}
\begin{figure}
        \centering
        \includegraphics[width=0.47\textwidth]{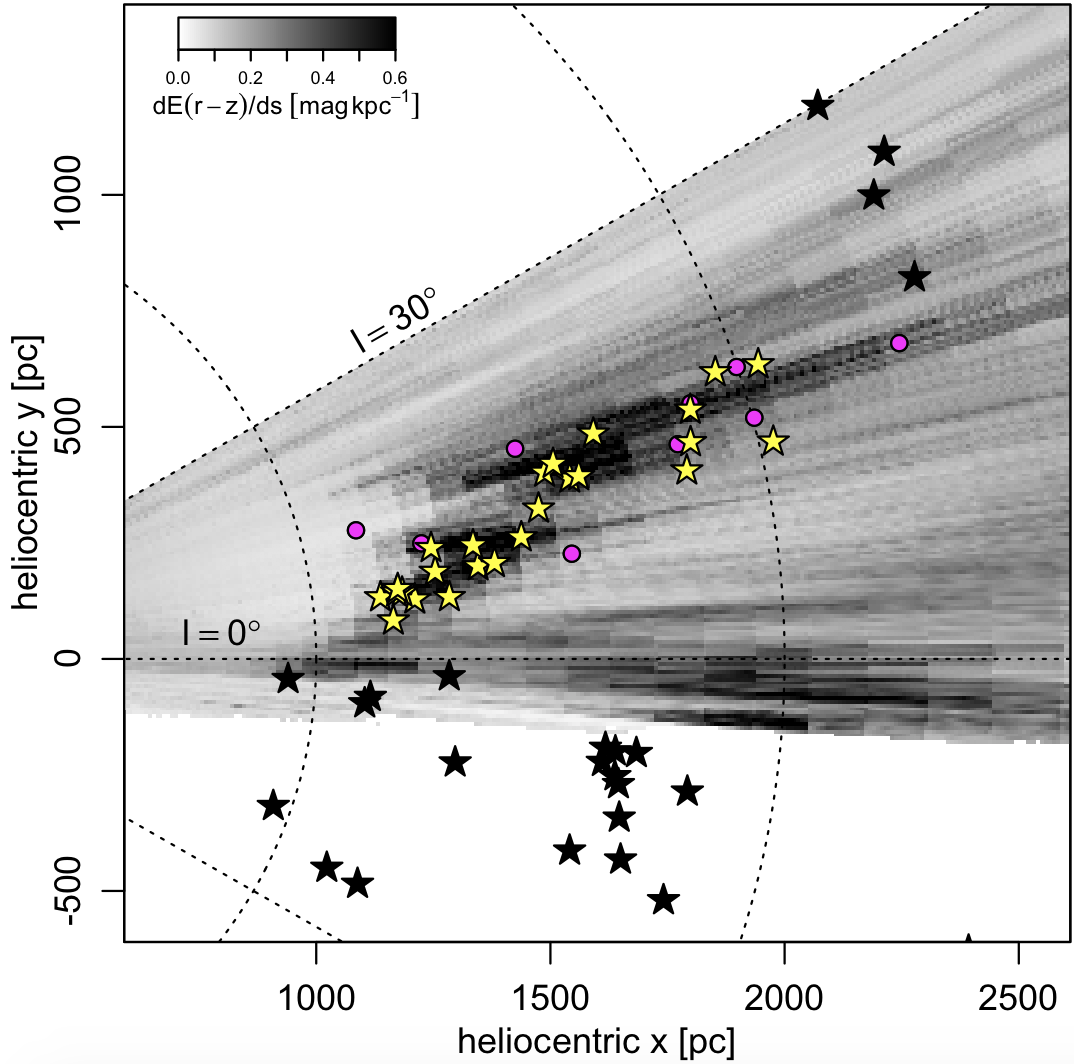}
    \caption{YSO groups and masers (same symbols as Fig.~\ref{fig:heliocentric_distribution}) overplotted on the 3D dust reddening map from \citet{Green2019}, integrated from $z=-300$~pc to 300~pc. The YSO groups follow a dust feature in the reddening map with a similar pitch angle. The dust map is truncated by the limits of the Pan-STARRS survey (bottom edge) and by our $|\ell|<30^\circ$ limits in this letter (top edge).
    }
  \label{fig:av}
\end{figure}

Three-dimensional dust extinction maps produced by \citet{Lallement2019}, \citet{Green2019}, and \citet{Zucker2020} reveal a filament-shaped region of high extinction that coincides with the structure formed by the YSO groups that we have examined in this paper (Fig.~\ref{fig:av}). The results from the extinction maps are effectively independent from our analysis of YSO group distances because they are based on photometry and \textit{Gaia} parallaxes of different sets of stars, with the 3D dust map based on main-sequence stars and our distances based on YSOs. The 3D dust maps support the view that the structure is a narrow linear feature with a width of $\approx 200$ pc and a length of $\approx 1$ kpc.  

\section{CO analysis}\label{sec:CO}

\begin{figure}
        \centering
        \includegraphics[width=0.46\textwidth]{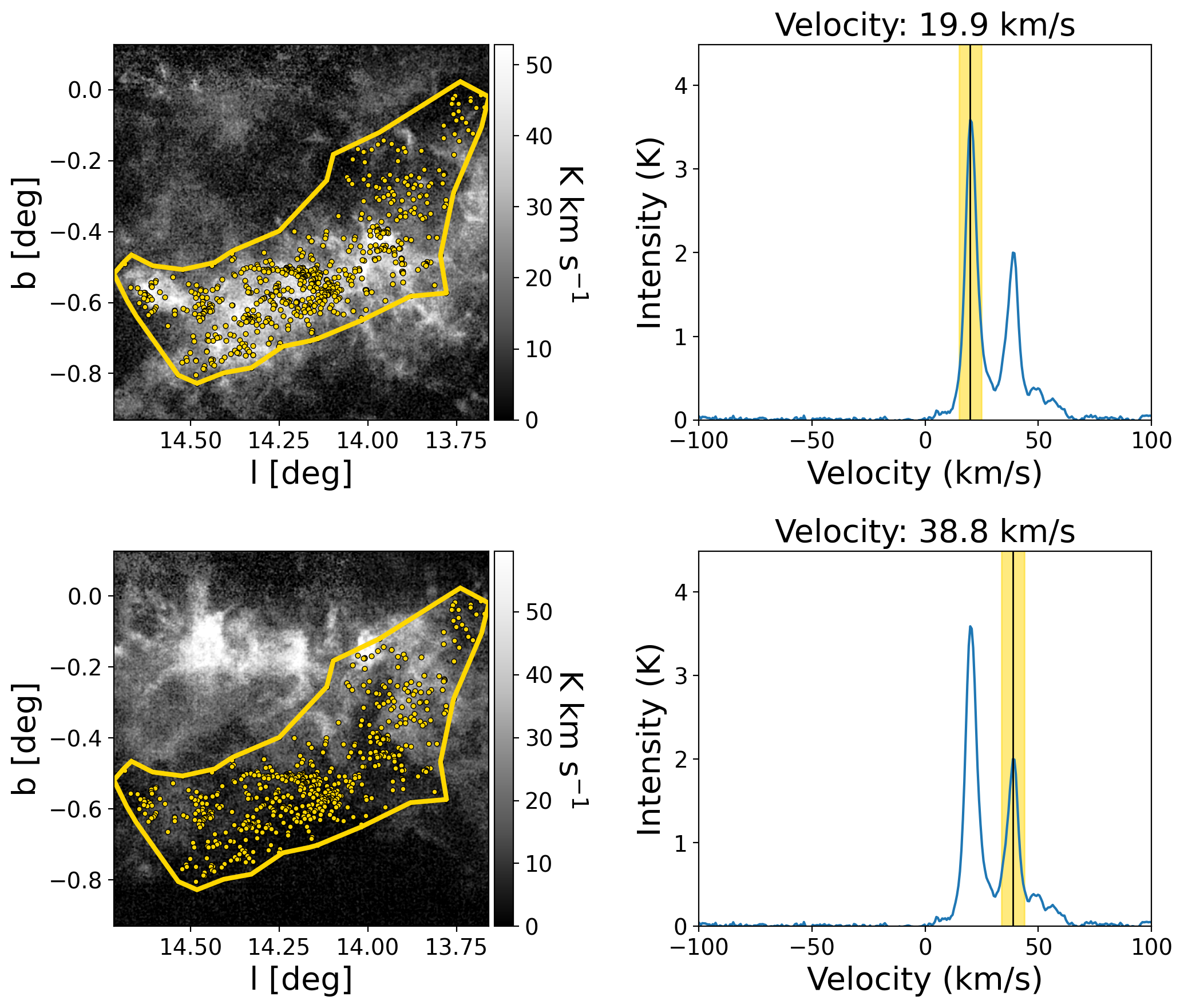}
    \caption{Plots of $^{13}$CO $J$=1--0 emission from the SEDIGISM survey \citep{Schuller2021} for G14.1-0.5 ($=$~M17 SWex). The left column is integrated emission over two different 10~km~s$^{-1}$ velocity ranges, and the right column is the CO spectrum, with the velocity range centered at 19.9 and 38.8 km~s$^{-1}$ (yellow shading). The grouped SPICY objects (yellow dots) are plotted on the integrated emission map for comparison with the spatial distribution of CO gas. Here the YSOs show the strongest spatial correlation with the lower velocity component at 20~km~s$^{-1}$.}
  \label{fig:co_individual}
\end{figure}

For each YSO group, we characterized the velocity structure of associated molecular gas using available CO spectral-line surveys. We favored the $^{13}{\rm CO}$ line whenever possible since it is optically thinner than $^{12}{\rm CO}$. We prioritized the use of the $^{13}{\rm CO}$ survey based on angular and spectral resolution, first checking whether a group exists in the SEDIGISM \citep{Schuller2021} footprint, followed by FUGIN \citep{Umemoto2017} and THrUMMS \citep{Barnes2015}. For YSO groups that lie outside the boundaries of existing $^{13}{\rm CO}$ surveys, we supplemented the $^{13}{\rm CO}$ data with $^{12}{\rm CO}$ data from the 1.2 m CfA CO survey \citep{Dame2001}, offering full coverage of the Galactic plane.

To extract a spectrum, we started by defining a concave hull on the plane of the sky using the stellar members of each group, as delineated by the yellow polygon on the left hand side of Fig. \ref{fig:co_individual} for YSO group G14.5-1.0. To do so, we employed the \texttt{Alpha Shape} Python package \citep{alphashape}, adopting a uniform alpha parameter of 5.0. We then identified the peak velocity component of the spectrum averaged over the area inside the concave hull. Since the YSO group is not necessarily associated with the highest intensity gas component, we employed a peak finding algorithm to identify all potentially relevant velocity maxima. To do so, we used the \texttt{find\_peaks} algorithm in the \texttt{SciPy} signal processing module, with the conservative criteria that any sub-peak must have a height of at least $25\%$ of the main peak, have a width of at least $2 \; \rm km \; s^{-1}$, and be at least $2 \; \rm km \; s^{-1}$ away from neighboring peaks. Then for each peak we created a custom zeroth moment map, integrated $\pm \rm 5 \; km \; s^{-1}$ from the peak velocity, which we contextualized in light of the spatial distribution of YSOs on the plane of the sky. The spectrum for G14.5-1.0 is shown on the right hand side of Fig.~\ref{fig:co_individual}. Group G14.5-1.0 has two velocity peaks, at $19.9 \rm \; km \; s^{-1}$ and $36.3 \rm \; km \; s^{-1}$, but only the lower velocity component shows strong spatial correlation with the YSO structure and group concave hull boundary on the plane of the sky.

We repeated the procedure shown in Fig.~\ref{fig:co_individual} for the remaining YSO groups, with three authors evaluating the spatial correspondence between the YSO groups and the zeroth moment maps integrated around each velocity peak. All groups associated with the structure show a plausible component at the velocity of the near Sagittarius arm, but many also show additional strong peaks at higher velocities, consistent with the near Scutum arm at larger distances (see Fig.~\ref{fig:COGAL_arm}). Given the significant confusion along the line of sight, we are making all the potential velocities for each group  publicly available, and the most plausible velocities are plotted in Fig.~\ref{fig:COGAL_arm}. Future targeted observations of YSOs in the structure obtained with high-resolution stellar spectroscopy should allow for more refined estimates of its velocity structure.

To transform to galactocentric coordinates (e.g., Fig.~\ref{fig:galactocentric}), we assumed velocity uncertainties of $\pm$5~km~s$^{-1}$ for unambiguous $v_\mathrm{lsr}$ solutions. This is similar to the peak widths observed for many clouds and is also the velocity range used for the integrated emission maps in Fig.~\ref{fig:co_individual}. For the six ambiguous cases, we picked the solution most consistent with the expected Sagittarius Arm velocity with an uncertainty equal to the difference between the maximum and minimum possible $v_\mathrm{lsr}$ solutions.

\end{appendix}

\begin{acknowledgements}
This work has made use of data from the European Space Agency (ESA) mission
{\it Gaia} (\url{https://www.cosmos.esa.int/gaia}), processed by the {\it Gaia}
Data Processing and Analysis Consortium (DPAC,
\url{https://www.cosmos.esa.int/web/gaia/dpac/consortium}). Funding for the DPAC
has been provided by national institutions, in particular the institutions
participating in the Gaia Multilateral Agreement. This work is based in part on archival data obtained with the {\it Spitzer} Space Telescope, which was operated by the Jet Propulsion Laboratory, California Institute of Technology under a contract with NASA. M.A.K.\ was partially supported by {\it Chandra} grant GO9-20002X. R.A.B.\ would like to acknowledge support from NASA grant NNX17AJ27G. AKM acknowledges the support from the Portuguese Funda\c c\~ao para a Ci\^encia e a Tecnologia (FCT) grants UID/FIS/00099/2019, PTDC/FIS-AST/31546/2017. We thank Butler Burton, Eve Ostriker, Tom Dame, and Debra Elmegreen for useful discussions, and we thank the referee for a thorough and thoughtful report on the original manuscript. The Cosmostatistics Initiative (COIN, \url{https://cosmostatistics-initiative.org/}) is an international network of researchers whose goal is to foster interdisciplinarity inspired by Astronomy.
\end{acknowledgements}


\begin{thebibliography}{}
 
 \bibitem[Avedisova(1989)]{Avedisova1989} Avedisova, V.~S.\ 1989, Astrophysics, 30, 83

 \bibitem[Alves et al.(2020)]{Alves2020} Alves, J., Zucker, C., Goodman, A.~A., et al.\ 2020,  \nat, 578, 237

 \bibitem[Balbus(1988)]{Balbus1988} Balbus, S.~A.\ 1988, \apj, 324, 60. doi:10.1086/165880

 \bibitem[Barnes et al.(2015)]{Barnes2015} Barnes, P.~J., Muller, E., Indermuehle, B., et al.\ 2015, \apj, 812, 6. doi:10.1088/0004-637X/812/1/6

 \bibitem[Bellock et al. (2021)]{alphashape} Bellock, K., Godber, N., Kahn, P. \ 2021, Zenodo. 10.5281/zenodo.4603211

  \bibitem[Benjamin et al.(2003)]{Benjamin2003} Benjamin, R.~A., Churchwell, E., Babler, B.~L., et al.\ 2003, \pasp, 115, 953
  
  \bibitem[Binder \& Povich(2018)]{Binder2018} Binder, B.~A. \& Povich, M.~S.\ 2018, \apj, 864, 136
  
  \bibitem[Bland-Hawthorn \& Gerhard(2016)]{Bland-Hawthorn_Gerhard2016} Bland-Hawthorn, J. \& Gerhard, O.\ 2016, \araa, 54, 529
   
\bibitem[Burton(1966)]{Burton1966} Burton, W.~B.\ 1966, \bain, 18, 247

\bibitem[Burton(1971)]{Burton1971} Burton, W.~B.\ 1971, \aap, 10, 76

\bibitem[Burton(1972)]{Burton1972} Burton, W.~B.\ 1972, \aap, 19, 51

\bibitem[Burton \& Bania(1974)]{Burton1974} Burton, W.~B. \& Bania, T.~M.\ 1974, \aap, 33, 425

\bibitem[Burton \& Shane(1970)]{Burton1970} Burton, W.~B. \& Shane, W.~W.\ 1970, The Spiral Structure of our Galaxy, 38, 397

   \bibitem[Calzetti et al.(2007)]{Calzetti2007} Calzetti, D., Kennicutt, R.~C., Engelbracht, C.~W., et al.\ 2007, \apj, 666, 870

   \bibitem[Campello et al.(2013)]{Campello2013} Campello, R.\ J., Moulavi, D., \& Sander, J.\ 2013, in
   Pacific-Asia conference on knowledge discovery and data
   mining, Springer, 160
 
  \bibitem[Cantat-Gaudin et al.(2020)]{Cantat-Gaudin2020} Cantat-Gaudin, T., Anders, F., Castro-Ginard, A., et al.\ 2020, \aap, 640, A1. doi:10.1051/0004-6361/202038192

   \bibitem[Churchwell et al.(2009)]{Churchwell2009} Churchwell, E., Babler, B.~L., Meade, M.~R., et al.\ 2009, \pasp, 121, 213

   \bibitem[Cleveland(1992)]{Cleveland1992} Cleveland, W. S., Grosse, E., \& Shyu, W.M.\ 1992, in Statistical Models in S, ed. Chambers, J. M.\ \& Hastie, T. J. (New York, NY: Wadsworth \& Brooks/Cole)

   \bibitem[Cohen et al.(1980)]{Cohen1980} Cohen, R.~S., Cong, H., Dame, T.~M., et al.\ 1980, \apjl, 239, L53

   \bibitem[Dame et al.(2001)]{Dame2001} Dame, T.~M., Hartmann, D., \& Thaddeus, P.\ 2001, \apj, 547, 792
 
  \bibitem[Dame et al.(1986)]{Dame1986} Dame, T.~M., Elmegreen, B.~G., Cohen, R.~S., et al.\ 1986, \apj, 305, 892
 
 \bibitem[Dobbs et al.(2006)]{Dobbs2006} Dobbs, C.~L., Bonnell, I.~A., \& Pringle, J.~E.\ 2006, \mnras, 371, 1663. doi:10.1111/j.1365-2966.2006.10794.x

  \bibitem[Downes et al.(1980)]{Downes1980} Downes, D., Wilson, T.~L., Bieging, J., et al.\ 1980, \aaps, 40, 379
 
  \bibitem[Elmegreen(1980)]{Elmegreen1980} Elmegreen, D.~M.\ 1980, \apj, 242, 528

\bibitem[Elmegreen(1985)]{Elmegreen1985} Elmegreen, D.~M.\ 1985, The Milky Way Galaxy, 106, 255

\bibitem[Eilers et al.(2019)]{Eilers2019} Eilers, A.-C., Hogg, D.~W., Rix, H.-W., et al.\ 2019, \apj, 871, 120

  \bibitem[Ewen \& Purcell(1951)]{Ewen1951} Ewen, H.~I. \& Purcell, E.~M.\ 1951, \nat, 168, 356

   \bibitem[Hogg et al.(2010)]{Hogg2010} Hogg, D.~W., Bovy, J., \& Lang, D.\ 2010, arXiv:1008.4686

 \bibitem[Hou \& Han(2015)]{Hou2015} Hou, L.~G. \& Han, J.~L.\ 2015, \mnras, 454, 626

   \bibitem[Gaia Collaboration et al.(2016)]{Gaia2016} Gaia Collaboration, Prusti, T., de Bruijne, J.~H.~J., et al.\ 2016, \aap, 595, A1. doi:10.1051/0004-6361/201629272

   \bibitem[Gaia Collaboration et al.(2018)]{Gaia2018} Gaia Collaboration, Brown, A.~G.~A., Vallenari, A., et al.\ 2018, \aap, 616, A1. doi:10.1051/0004-6361/201833051

  \bibitem[Gaia Collaboration et al.(2021)]{Gaia2020} Gaia Collaboration, Brown, A.~G.~A., Vallenari, A., et al.\ 2021, \aap, 649, A1

   \bibitem[Gelman \& Rubin(1992)]{Gelman1992} Gelman, A. \& Rubin, D.~B.\ 1992, Statistical Science, 7, 457. doi:10.1214/ss/1177011136

   \bibitem[Georgelin \& Georgelin(1976)]{Georgelin1976} Georgelin, Y.~M. \& Georgelin, Y.~P.\ 1976, \aap, 49, 57

   \bibitem[Gerasimenko(1993)]{Gerasimenko1993} Gerasimenko, T.~P.\ 1993, \azh, 70, 953

    \bibitem[Getman et al.(2014)]{Getman2014} Getman, K.~V., Feigelson, E.~D., Kuhn, M.~A., et al.\ 2014, \apj, 787, 108

\bibitem[Gingerich(1985)]{Gingerich1985} Gingerich, O.\ 1985, The Milky Way Galaxy: IAU Symposium 106, 59

\bibitem[Gravity Collaboration et al.(2019)]{GravityCollaboration2019} Gravity Collaboration, Abuter, R., Amorim, A., et al.\ 2019, \aap, 625, L10
   
   \bibitem[Green et al.(2019)]{Green2019} Green, G.~M., Schlafly, E., Zucker, C., et al.\ 2019, \apj, 887, 93. doi:10.3847/1538-4357/ab5362
  
  \bibitem[Gvaramadze et al.(2014)]{Gvaramadze2014} Gvaramadze, V.~V., Menten, K.~M., Kniazev, A.~Y., et al.\ 2014, \mnras, 437, 843. doi:10.1093/mnras/stt1943
  
   \bibitem[Kerr(1969)]{Kerr1969} Kerr, F.~J.\ 1969, \araa, 7, 39

  \bibitem[Kim \& Ostriker(2002)]{Kim2002} Kim, W.-T. \& Ostriker, E.~C.\ 2002, \apj, 570, 132. doi:10.1086/339352

   \bibitem[Kim et al.(2020)]{Kim2020} Kim, W.-T., Kim, C.-G., \& Ostriker, E.~C.\ 2020, \apj, 898, 35

   \bibitem[Kuhn et al.(2020)]{spicyI} Kuhn, M. A., de Souza, R. S., Krone-Martins, A., Castro-Ginard, A., Ishida, E. E. O., Povich, M. S., Hillenbrand, L. A. 2020,
      ApJS, submitted

   \bibitem[Kuhn \& Hillenbrand(2020)]{KuhnHillenbrand2020} Kuhn, M.~A. \& Hillenbrand, L.~A.\ 2020, Research Notes of the American Astronomical Society, 4, 224

   \bibitem[Kuhn et al.(2019)]{Kuhn2019} Kuhn, M.~A., Hillenbrand, L.~A., Sills, A., et al.\ 2019, \apj, 870, 32

   \bibitem[Kuhn \& Feigelson(2019)]{HoMA} Kuhn, M.~A. \& Feigelson, E.~D.\ 2019, in Handbook of Mixture Analysis, ed.\ G.\ Celeux, S.\ Fr\"uwirth-Schnatter, \& C.\ P.\ Robert (Chapman \& Hall/CRC), 463

   \bibitem[Kwee et al.(1954)]{Kwee1954} Kwee, K.~K., Muller, C.~A., \& Westerhout, G.\ 1954, \bain, 12, 211

  \bibitem[Lallement et al.(2019)]{Lallement2019} Lallement, R., Babusiaux, C., Vergely, J.~L., et al.\ 2019, \aap, 625, A135

  \bibitem[La Vigne et al.(2006)]{LaVigne2006} La Vigne, M.~A., Vogel, S.~N., \& Ostriker, E.~C.\ 2006, \apj, 650, 818

  \bibitem[Lindegren et al.(2021a)]{Lindegren2020_zp} Lindegren, L., Bastian, U., Biermann, M., et al.\ 2021, \aap, 649, A4

  \bibitem[Lindegren et al.(2021b)]{Lindegren2020_as} Lindegren, L., Klioner, S.~A., Hern{\'a}ndez, J., et al.\ 2021, \aap, 649, A2
    
    \bibitem[Mills(1959)]{Mills1959} Mills, B.~Y.\ 1959, \pasp, 71, 267
    
   \bibitem[Morgan et al.(1955)]{Morgan1955} Morgan, W.~W., Code, A.~D., \& Whitford, A.~E.\ 1955, \apjs, 2, 41

   \bibitem[Morgan et al.(1953)]{Morgan1953} Morgan, W.~W., Whitford, A.~E., \& Code, A.~D.\ 1953, \apj, 118, 318
    
   \bibitem[Muller \& Oort(1951)]{Muller1951} Muller, C.~A. \& Oort, J.~H.\ 1951, \nat, 168, 357
    
  \bibitem[Pantaleoni Gonz{\'a}lez et al.(2021)]{PantaleoniGonzalez2021} Pantaleoni Gonz{\'a}lez, M., Ma{\'\i}z Apell{\'a}niz, J., Barb{\'a}, R.~H., et al.\ 2021, \mnras, 504, 2968

 \bibitem[Pawsey(1951)]{Pawsey1951} Pawsey, J.L.\ 1951, \nat, 168, 358

   \bibitem[Plummer(2019)]{Plummer2019} Plummer, M.\ 2019, rjags: Bayesian Graphical Models using MCMC, \url{https://CRAN.R-project.org/package=rjags}

  \bibitem[Poggio et al.(2021)]{Poggio2021} Poggio, E., Drimmel, R., Cantat-Gaudin, T., et al.\ 2021, arXiv:2103.01970
   
   \bibitem[Prisinzano et al.(2019)]{Prisinzano2019} Prisinzano, L., Damiani, F., Kalari, V., et al.\ 2019, \aap, 623, A159. doi:10.1051/0004-6361/201834870
      
   \bibitem[R Core Team(2019)]{RCore} R Core Team (2019). R: A language and environment for statistical computing. R
  Foundation for Statistical Computing, Vienna, Austria
   
   \bibitem[Reid et al.(2016)]{Reid2016} Reid, M.~J., Dame, T.~M., Menten, K.~M., et al.\ 2016, \apj, 823, 77

   \bibitem[Reid et al.(2019)]{Reid2019} Reid, M.~J., Menten, K.~M., Brunthaler, A., et al.\ 2019, \apj, 885, 131

   \bibitem[Reid et al.(2014)]{Reid2014} Reid, M.~J., Menten, K.~M., Brunthaler, A., et al.\ 2014, \apj, 783, 130
   
   \bibitem[Rho et al.(2008)]{Rho2008} Rho, J., Lefloch, B., Reach, W.~T., et al.\ 2008, Handbook of Star Forming Regions, Volume II, 509

    \bibitem[Sandage(1961)]{Sandage1961} Sandage, A.\ 1961, Washington: Carnegie Institution, 1961
  
    \bibitem[Schmidt(1957)]{Schmidt1957} Schmidt, M.\ 1957, \bain, 13, 247
  
    \bibitem[Schuller et al.(2021)]{Schuller2021} Schuller, F., Urquhart, J.~S., Csengeri, T., et al.\ 2021, \mnras, 500, 3064

   \bibitem[Shetty \& Ostriker(2006)]{Shetty2006} Shetty, R. \& Ostriker, E.~C.\ 2006, \apj, 647, 997. doi:10.1086/505594

   \bibitem[Simonson(1970)]{Simonson1970} Simonson, S.~C.\ 1970, \aap, 9, 163

   \bibitem[Su (2020)]{Su2020} Su, Y.-S. 2020, R2jags: Using R to Run JAGS, \url{https://CRAN.R-project.org/package=R2jags}

   \bibitem[Taylor \& Cordes(1993)]{TaylorCordes93} Taylor, J.~H. \& Cordes, J.~M.\ 1993, \apj, 411, 674 

   \bibitem[Umemoto et al.(2017)]{Umemoto2017} Umemoto, T., Minamidani, T., Kuno, N., et al.\ 2017, \pasj, 69, 78
   
   \bibitem[van de Hulst et al.(1954)]{vanDeHulst1954} van de Hulst, H.~C., Muller, C.~A., \& Oort, J.~H.\ 1954, \bain, 12, 117
   
   \bibitem[van der Kruit \& Freeman(2011)]{vanDerKruit2011} van der Kruit, P.~C. \& Freeman, K.~C.\ 2011, \araa, 49, 301
   
   \bibitem[Venables \& Ripley(2002)]{VenablesRipley2002} Venables \& Ripley 2002, Modern Applied Statistics with S, {New York, NY: Springer}

   \bibitem[VERA Collaboration et al.(2020)]{Hirota_2020} VERA Collaboration, Hirota, T., Nagayama, T., et al.\ 2020, \pasj, 72, 50. doi:10.1093/pasj/psaa018
   
   \bibitem[Wada \& Koda(2004)]{Wada2004} Wada, K. \& Koda, J.\ 2004, \mnras, 349, 270

   \bibitem[Westerhout(1958)]{Westerhout1958} Westerhout, G.\ 1958, \bain, 14, 215
   
   \bibitem[Weaver(1970)]{Weaver1970} Weaver, H.\ 1970, The Spiral Structure of our Galaxy, 38, 126

   \bibitem[Wu et al.(2014)]{Wu2014} Wu, Y.~W., Sato, M., Reid, M.~J., et al.\ 2014, \aap, 566, A17
   
   \bibitem[Xu et al.(2021)]{Xu2021} Xu, Y., Hou, L.~G., Bian, S.~B., et al.\ 2021, \aap, 645, L8. doi:10.1051/0004-6361/202040103
   
   \bibitem[Zari et al.(2021)]{Zari2021} Zari, E., Rix, H.-W., Frankel, N., et al.\ 2021, \aap, 650, A112

   \bibitem[Zucker et al.(2020)]{Zucker2020} Zucker, C., Speagle, J.~S., Schlafly, E.~F., et al.\ 2020, \aap, 633, A51

\end{thebibliography}
\end{document}